\documentclass[a4paper,twoside,11pt]{article} % to use for all kind of papers but AACA

\usepackage[english]{babel} %
\usepackage{lmodern}
\usepackage[T1]{fontenc}
\usepackage[latin9]{inputenc}	% Così le "è" vengono trasformate in \`e etc. etc.
\newlength{\defaultparindent}
\usepackage{latexsym}
\usepackage{amsmath}
\usepackage{amsfonts}
\usepackage{amsthm}
\usepackage{bbold}
\usepackage{multirow}
\usepackage{hyperref}
\hypersetup{colorlinks=true,citecolor=green,linkcolor= blue}
\usepackage{todonotes} % more appealing than \marginpar ??
\usepackage{soul} % to get overstriked text

\usepackage{algorithm} % to write algorithms, suggested by Renato on May 11, 2017
\usepackage{algpseudocode}

\usepackage[arXiv]{optional} % Possible values of the option fixing the format: x, std, arXiv, JMP, JOPA, AACA and also note control: margin_notes, final_notes
\def\mynote{\todo} % \marginpar or \todo
\opt{AACA}{% only for submission to AACA -> \cal font is not defined....
\def\cal{\mathcal}
}

\opt{x,AACA}{% in all cases - but arXiv, std, JMP & JOPA - include my file of standard defs: mbh_defs_TeX
\input{mbh_defs_TeX}% My standard TeX definitions, original in Dictionaries & Gloassaries folder
}

\opt{arXiv,std,JMP,JOPA}{% only for arXiv, std, JMP & JOPA we need the definitions included in mbh_defs_TeX (that contains original versions)

\newtheorem{MS_theorem}{Theorem}
\newtheorem{MS_lemma}{Lemma}
\newtheorem{MS_Proposition}{Proposition}
\newtheorem{MS_Corollary}[MS_Proposition]{Corollary}
\newtheorem{MS_Algorithm}{Algorithm}
\def\myconjugate#1{\overline{#1}} % this way conjugate definitions can be easily changed (\bar did not work fine on multiple chars)

\def\eg{e.g.\ }
\def\myisom{\cong} % this way isomorphism symbol definition can be easily changed; possiblities: \cong \simeq
\newcommand{\R}{\ensuremath{\mathbb{R}}} % good; MB XII 2005; needs \usepackage{bbold}
\newcommand{\Identity}{\ensuremath{\mathbb{1}}} % good; MB XII 2005; needs \usepackage{bbold}
\def\my_span#1{\mbox{Span}\left(#1\right)} % changed from 'span' since it interfered with \multicolumn{} MB X 2009
\def\dotinformula{\;\; \mathrm{.}} % defines space + a full stop (in \rm font) to be placed at the end of a formula

\newcommand{\comm}[2]{\ensuremath{\left[ #1, #2 \right]}}
\newcommand{\anticomm}[2]{\ensuremath{\left\{ #1, #2 \right\}}} % or \left[...\right]^+
\newcommand{\myClg}[3]{\ensuremath{{{\cal C}\ell} {\left( #3 \right)}}}	% this is a possible def of myCl that uses only the scalar product g

}

\opt{JMP}{% only for submission to JMP -> double line spacing
}

\def\h_eigen{\eta}
\def\g_eigen{\theta}
\def\mygen{e} % this way the definition of the generators of the algebra can be easily changed
\def\myprimidemp{\mathbb{p}} % this way the definition of the primitive idempotents can be easily changed (beware: not all fonts in math mode have lower characters, e.g. \cal)

\def\SAT{\ensuremath{\mbox{SAT}}}
\def\bigO#1{\ensuremath{\mathcal{O}\left(#1\right)}}
\def\literal{\ensuremath{\rho}}
\def\myAbelsa{\mathcal{P}} % this way the definition of the Abelian subalgebra of Cl(m,m) can be easily changed
\def\mysetS{\mathcal{I}} % this way the definition of the set S can be easily changed
\def\myBooleanS{{\cal{S}}} % this way the definition of the a Boolean SAT problem S can be easily changed
\def\myGenVarS{{\mathbb{S}}} % this way the definition of the generalized variables S can be easily changed

\begin{document}

\opt{x,std,arXiv,JMP,JOPA}{% in all cases - but AACA
\title{{\bf The Boolean SATisfiability Problem in Clifford algebra} %\\(temporary title)
	}

\author{\\
	\bf{Marco Budinich}%
%
%\footnote{on leave of absence from: University of Trieste, Trieste, Italy}%
%
\\
%	ICTP and INFN, Trieste, Italy\\
	University of Trieste and INFN, Trieste, Italy\\
	\texttt{mbh@ts.infn.it}\\
%	\texttt{http://www.ts.infn.it/\~{ }mbh/MBHgeneral.html}\\
%
%\\	Very preliminary - restricted circulation (FYEO)
%
%
%	Submitted on March 17, 2016
%	Submitted to: {\em Journal of Mathematical Physics} on September 29, 2017
%		keywords of this submission: Spinor, Satisfiability problem, Clifford algebra
%	Submitted to: {\em Communications in Mathematical Physics} on March 14, 2016.\\
%	Submitted to: {\em Journal of Physics A: Mathematical and Theoretical} \\on October 20, 2017
%	Submitted to: {\em SIAM Journal on Discrete Mathematics} on March 30, 2018
%\\	Re-resubmitted to: {\em Advances in Applied Clifford Algebras} on June 12, 2017
%	Published in: {\em Advances in Applied Clifford Algebras}, 2015\\
%	{\small DOI:10.1007/s00006-015-0547-8}
%	Resubmitted to: {\em Journal of Mathematical Physics} on June 27, 2016\\
%	{\tiny First submission: March 30, 2016}
%	To appear in: {\em Journal of Mathematical Physics}\\
%	{\tiny Submitted on March 30 and June 27, 2016; accepted July 12, 2016}
%	{\em Journal of Mathematical Physics} {\bf 57} (2016) DOI: 10.1063/1.4959531\\
%	\\ May 29, 2017, submitted
	}
\date{ \today }
%\date{April 21, 2017}
%\date{ } % to hide the date this line must be present (believe it or not...)
\maketitle
}

\opt{AACA}{% only for AACA
\title[On Spinors of Zero Nullity]{On Spinors of Zero Nullity}

\author{Marco Budinich}
\address{Dipartimento di Fisica\\
	Università di Trieste \& INFN\\
	Via Valerio 2, I - 34127 Trieste, Italy}
\email{mbh@ts.infn.it}
}

\begin{abstract}
We present a formulation of the Boolean Satisfiability Problem in spinor language that allows to give a necessary and sufficient condition for unsatisfiability. With this result we outline an algorithm to test for unsatisfiability with possibly interesting theoretical properties.
\end{abstract}

%\opt{x,std,arXiv,JMP,JOPA}{% in all cases - but AACA
%\noindent{\bf Keywords:} {Spinors, discrete transformations, Clifford algebra.}
%}

\opt{AACA}{% only for AACA
\maketitle
}

\section{Introduction}
\label{Introduction}
\opt{margin_notes}{\mynote{mbh.note: for paper material see log pp. 689 ff.}}%
In 1913 {\'{E}}lie Cartan introduced spinors \cite{Cartan_1913} and, after more than a century, this field still yields rich harvests.
% for possible incipits see file: incipit_spinor_papers
Spinors were later thoroughly investigated by Claude Chevalley \cite{Chevalley_1954} in the mathematical frame of Clifford algebras where they were identified as elements of Minimal Left Ideals of the algebra.

%In this paper we write the famous Boolean Satisfiability Problem (SAT) in spinor language and exploit the properties of Clifford algebra to arrive to an algebraic necessary and sufficient condition for unsatisfiability. Beyond the well established statistical mechanics formulation of SAT \cite{Braunstein_2005} this paper offers another connection between SAT and physics exploiting the properties of the algebra of fermions.

%The famous Boolean Satisfiability Problem (SAT) have been already dealt with by physics in the language of statistical mechanics \cite{Braunstein_2005}, here we offer another link between SAT and physics based on the algebra of fermions: we will exploit the properties of Clifford algebra to arrive to an algebraic necessary and sufficient condition for unsatisfiability.

In this paper we establish a link between the famous Boolean Satisfiability Problem (\SAT{}) and the algebra of fermions: we exploit the properties of Clifford algebra to give an algebraic formulation of \SAT{} that yields a necessary and sufficient condition for unsatisfiability. This is not the first time that physics meets \SAT{}: also statistical mechanics methods applied to \SAT{} have obtained considerable successes \cite{Braunstein_2005}.

In section~\ref{SAT_basics} we succinctly resume basic properties of \SAT{} and in section~\ref{Boolean_formulation} we introduce an Abelian subalgebra of Clifford algebra and show that any Boolean expression is neatly represented by an idempotent of this subalgebra. In following section~\ref{SAT_reformulation} we apply this method to transfer \SAT{} problems in Clifford algebra and in section~\ref{unSAT_condition} we exploit this formulation to obtain, with a basic theorem of Clifford algebra, a necessary and sufficient condition for a \SAT{} problem to be unsatisfiable. In final section~\ref{unSAT_algorithm} we use this result to outline an actual algorithm for \SAT{} problems that turns out to be a relaxation of the standard Davis Putnam recursive algorithm \cite{Davis_Weyuker_1983}.

% and \ref{SAT_reformulation} we present an equivalent representation of \SAT{} in an Abelian subalgebra of Clifford algebra and we show that more in general any Boolean expression can be neatly represented by an idempotent of this subalgebra. In the following 
For the convenience of the reader we tried to make this paper as elementary and self-contained as possible.

\section{Satisfiability problem}
\label{SAT_basics}
The Boolean Satisfiability Problem \cite[Section~7.2.2.2]{Knuth_2015} asks for an assignment of $n$ Boolean variables $x_i \in \{\mathrm{T}, \mathrm{F}\}$ (true, false), that satisfies a given Boolean formula expressed in Conjunctive Normal Form (CNF) \eg
$$
(x_1 \lor \lnot x_2) \land (x_2 \lor x_3) \land ( \lnot x_1 \lor \lnot x_3) \land ( \lnot x_1 \lor \lnot x_2 \lor x_3) \land (x_1 \lor x_2 \lor \lnot x_3)
$$
as a logical AND ($\land$) of $m$ \emph{clauses} ${\cal C}_j$, the expressions in parenthesis, each clause being composed by the logical OR ($\lor$) of $k$ or less Boolean variables possibly preceded by logical NOT ($\lnot$); in this example $n = 3, m = 5, k = 3$. A solution of the problem is either an assignment of $x_i$ that makes the expression $\mathrm{T}$ or a proof that such an assignment does not exist.

\SAT{} was the first combinatorial problem proven to be NP-complete \cite{Cook_1971}; in particular while the case of $k = 3$, 3\SAT, can be solved only in a time that grows exponentially with $n$, $2\SAT$ and $1\SAT$ problems can be solved in polynomial time (that is \emph{fast}).

A $1\SAT$ problem is just a logical AND of $m$ Boolean variables. For both assignments of $x_i$, $x_i \land \lnot {x}_i \equiv \mathrm{F}$ and thus the presence of a variable together with its logical complement is a necessary and sufficient condition for making a $1\SAT$ formula unsatisfiable. We use $\equiv$ for logical equivalence, namely that for all possible values taken by the variables the two expressions are equal, to mark the difference with algebraic equality $=$. We can interpret a satisfiable $1\SAT$ formula as an assignment of variables since there is only one assignment that makes it $\mathrm{T}$ and that can be read scanning the formula; in the sequel we will freely use $1\SAT$ formulas for assignments.

Using the distributive properties of the logical operators $\lor, \land$ any given $k\SAT$ formula expands in a logical OR of up to $k^m$ $1\SAT$ terms this being just an upper bound since terms containing $x_i \land \lnot {x}_i$ can be omitted. The final expanded expression is easily simplified and reordered exploiting the commutativity of the logical operators $\lor, \land$ and the properties $x_i \land x_i \equiv x_i \lor x_i \equiv x_i$. It is easy to see that all ``surviving'' terms of this expansion, the Disjunctive Normal Form (DNF), are $1\SAT$ terms, each of them representing an assignment that satisfies the problem. On the contrary if the DNF is empty, as happens for the example given above, this is a proof that no assignment can make the formula $\mathrm{T}$: the problem is unsatisfiable.

Expansion to DNF is a terrible algorithm for \SAT{}: first of all the method is an overkill since it produces all possible solutions whereas one is enough for \SAT{}; in second place this brute force approach gives a running time proportional to the number of expansion terms \bigO{(k^{\frac{m}{n}})^n} whereas the present best $\SAT$ solvers \cite{PaturiPudlakSaksZane_2005} run in \bigO{1.307^n}. On the other hand DNF will assume a precise meaning in Clifford algebra.

We will write a \SAT{} problem $\myBooleanS$ in a more concise form as
\begin{equation}
\label{formula_SAT_std}
\myBooleanS \equiv (\literal_i \lor \literal_j \lor \cdots \lor \literal_h) (\literal_l \lor \literal_o \lor \cdots \lor \literal_p) \cdots (\literal_t \lor \literal_u \lor \cdots \lor \literal_z) \quad \literal_i \in \{x_i, \myconjugate{x}_i \}
\end{equation}
where $\literal_i$ is a \emph{literal} that stands for Boolean variable $x_i$ or its complement, $\myconjugate{x}_i$ for short, and the ordinary product stands for logical AND.

\section{Boolean expressions in Clifford algebra}
\label{Boolean_formulation}
To tackle Boolean expressions in Clifford algebra we exploit the known fact that in any associative, unital, algebra every family of commuting, orthogonal, idempotents generates a Boolean algebra. This will allow us to put \SAT{} problems in algebraic form. We proceed step by step along this path with \SAT{} problems in mind.

Given a Boolean expression with $n$ Boolean variables $x_i$ we will represent it with an idempotent of the Clifford algebra $\myClg{}{}{\R^{n,n}}$ that is isomorphic to the algebra of real matrices $\R(2^n)$. This algebra is more easily manipulated exploiting the properties of its Extended Fock Basis (EFB, see \cite{Budinich_2016} and references therein) with which any algebra element is a linear superposition of simple spinors. The $2 n$ generators of the algebra $\mygen_{i}$ form an orthonormal basis of the neutral vector space $V = \R^{n,n}$ with \eg
\begin{equation}
\label{formula_generators}
\mygen_i \mygen_j + \mygen_j \mygen_i := \anticomm{\mygen_i}{\mygen_j} = 2 \delta_{i j} (-1)^{i+1} \qquad i,j = 1,2, \ldots, 2 n
\end{equation}
while the Witt, or null, basis of $V$ is:
\begin{equation}
\label{formula_Witt_basis}
\left\{ \begin{array}{l l l}
p_{i} & = & \frac{1}{2} \left( \mygen_{2i-1} + \mygen_{2i} \right) \\
q_{i} & = & \frac{1}{2} \left( \mygen_{2i-1} - \mygen_{2i} \right)
\end{array} \right.
%\Rightarrow
%\left\{\begin{array}{l l l}
%\mygen_{2i-1} & = & p_{i} + q_{i} \\
%\mygen_{2i} & = & p_{i} - q_{i}
%\end{array} \right.
\quad i = 1,2, \ldots, n
\end{equation}
that, with $\mygen_{i} \mygen_{j} = - \mygen_{j} \mygen_{i}$, gives
\begin{equation}
\label{formula_Witt_basis_properties}
\anticomm{p_{i}}{p_{j}} = \anticomm{q_{i}}{q_{j}} = 0
\qquad
\anticomm{p_{i}}{q_{j}} = \delta_{i j}
\end{equation}
showing that all $p_i, q_i$ are mutually orthogonal, also to themselves, that implies $p_i^2 = q_i^2 = 0$, at the origin of the name ``null'' given to these vectors.
\opt{margin_notes}{\mynote{mbh.note: commented there is a shorter form}}%
%
%Simple spinors forming EFB are products of $n$ or more null vectors (\ref{formula_Witt_basis}).
The $2^{2 n}$ simple spinors forming EFB are given by all possible sequences
\begin{equation}
\label{EFB_def}
\psi_1 \psi_2 \cdots \psi_n \qquad \psi_i \in \{ q_i p_i, p_i q_i, p_i, q_i \} \qquad i = 1,\ldots,n \dotinformula
\end{equation}
Since $\mygen_{2 i - 1} \mygen_{2 i} = q_i p_i - p_i q_i := \comm{q_i}{p_i}$ in EFB the identity $\Identity$ and the volume element $\omega$ (scalar and pseudoscalar) assume similar expressions \cite{Budinich_2016}:
\begin{equation}
\label{identity_omega}
\begin{array}{l l l}
\Identity & := & \anticomm{q_1}{p_1} \anticomm{q_2}{p_2} \cdots \anticomm{q_n}{p_n} \\
\omega & := & \mygen_1 \mygen_2 \cdots \mygen_{2 n} = \comm{q_1}{p_1} \comm{q_2}{p_2} \cdots \comm{q_n}{p_n} \dotinformula
\end{array}
\end{equation}

With (\ref{formula_Witt_basis_properties}) and $q_i p_i + p_i q_i = 1$ we easily obtain
$$
q_i p_i \, q_i p_i = q_i p_i \quad p_i q_i \, p_i q_i = p_i q_i \quad q_i p_i \, p_i q_i = p_i q_i \, q_i p_i = 0 \quad q_i p_i \, q_j p_j = q_j p_j \, q_i p_i
$$
that shows that $q_i p_i$ and $p_i q_i$ are part of a family of orthogonal, commuting, idempotents.
% and, with the associations:
%$$
%\literal_i \to q_i p_i \qquad \myconjugate{\literal}_i \to p_i q_i \qquad \mathrm{F} \to 0
%$$
%and mapping the logical AND to Clifford product, we recognize in previous relations the Boolean relations
%$$
%\literal_i \land \myconjugate{\literal}_i \equiv \myconjugate{\literal}_i \land \literal_i \equiv \mathrm{F} \quad \literal_i \land \literal_i \equiv \literal_i \quad \myconjugate{\literal}_i \land \myconjugate{\literal}_i \equiv \myconjugate{\literal}_i \quad \literal_i \land \literal_j \equiv \literal_j \land \literal_i \dotinformula
%$$
%From now on we will use $\literal_i$ and $\myconjugate{\literal}_i$ also in $\myClg{}{}{\R^{n,n}}$ meaning respectively $q_i p_i$ and $p_i q_i$ and Clifford product will stand for logical AND.
%
%To lift these promising relations to a full Boolean algebra we must add further structure.
Since $\myClg{}{}{\R^{n,n}}$ is a simple algebra, the unit element of the algebra is the sum of $2^n$ primitive (indecomposable) idempotents $\myprimidemp_i$
\begin{equation}
\label{formula_identity_def}
\Identity = \sum_{i = 1}^{2^n} \myprimidemp_i = \prod_{j = 1}^{n} \anticomm{q_j}{p_j}
\end{equation}
where the second form, a product of $n$ anticommutators, is its expression in EFB (\ref{identity_omega}). The full expansion of these anticommutators contains $2^n$ terms each term being one of the primitive idempotents \emph{and} a simple spinor (\ref{EFB_def}). We remind the standard properties of the primitive idempotents
\opt{margin_notes}{\mynote{mbh.note: by referee 2 primitive idempotents are Boolean \emph{atoms} (p. 2)}}%
\begin{equation}
\label{formula_primitive_idempotents}
\myprimidemp_i^2 = \myprimidemp_i \quad (\Identity - \myprimidemp_i)^2 = \Identity - \myprimidemp_i \quad \myprimidemp_i (\Identity - \myprimidemp_i) = 0 \quad \myprimidemp_i \myprimidemp_j = \delta_{i j} \myprimidemp_i \dotinformula
\end{equation}
Let $\myAbelsa \subset \myClg{}{}{\R^{n,n}}$ be the even, Abelian, subalgebra given by the linear space spanned by this family of $2^n$ primitive idempotents $\myprimidemp_i$. In the isomorphic matrix algebra $\R(2^n)$ $\myAbelsa$ is usually, but not necessarily, the subalgebra of diagonal matrices and in this case the primitive idempotents are the $2^n$ diagonal matrices with a single $1$ on the diagonal. Subalgebra $\myAbelsa$ contains a subset in which field coefficients are binary, namely
\begin{equation}
\label{formula_S_def}
\mysetS := \left\{ \sum_{i = 1}^{2^n} \delta_i \myprimidemp_i : \delta_i \in \left\{0, 1 \right\} \right\} \subset \myAbelsa
\end{equation}
that is closed under multiplication but not under addition and is thus not even a subspace. With primitive idempotent properties (\ref{formula_primitive_idempotents}) we can prove
\opt{margin_notes}{\mynote{mbh.note: the proof is commented. Is in general any idempotent a sum of primitive idempotents ?}}%
\begin{MS_Proposition}
\label{S_properties}
Given $s \in \myAbelsa$ then $s \in \mysetS$ if and only if $s$ is an idempotent, $s^2 = s$
\end{MS_Proposition}
%
%\begin{proof}
%Let $s \in \mysetS$, by primitive idempotent properties (\ref{formula_primitive_idempotents}) follows immediately that $s^2 = s$. Conversely let $s \in \myAbelsa$ be such that $s^2 = s$; let $s = \sum_{i = 1}^{2^n} \gamma_i \myprimidemp_i$, by primitive idempotent properties (\ref{formula_primitive_idempotents}) it follows that $s^2 = \sum_{i = 1}^{2^n} \gamma_i^2 \myprimidemp_i$ thus the hypothesis implies that $\gamma_i^2 = \gamma_i$ that in $\R$ holds only if $\gamma_i \in \{0, 1\}$.
%\end{proof}
%
\noindent and thus $\mysetS$ is the set of the idempotents, in general not primitive; a simple consequence is that for any $s \in \mysetS$ also $(\Identity - s) \in \mysetS$. $\mysetS$ properties are easier to grasp observing that $\delta_i \in \left\{0, 1 \right\}$, the only idempotents of $\R$.

We remark that the $2^n$ primitive idempotents $\{ \myprimidemp_i \}$ form an orthonormal basis%
\footnote{not strictly a basis since $\myprimidemp_i^2 = \myprimidemp_i$ but the set has all the other properties of an orthonormal basis}%
{} of $\myAbelsa$ as a linear space and thus uniquely identify all elements of $\myAbelsa$ and thus of $\mysetS$ (\ref{formula_S_def}) moreover by (\ref{formula_primitive_idempotents}) the $i$-th component of $s \in \myAbelsa$ is $\myprimidemp_i s$.
\opt{margin_notes}{\mynote{mbh.note: they are not \emph{exactly} an o.n. basis since $\myprimidemp_i^2 = \myprimidemp_i$ and not $\Identity$}}%

We are now ready to give the rules to replace Boolean elements with algebraic ones (using $\myconjugate{\literal}_i$ for $\lnot \literal_i$)
\begin{equation}
\label{Boolean_substitutions}
\begin{array}{lll}
\mathrm{F} & \to & 0 \\
\mathrm{T} & \to & \Identity \\
\literal_i & \to & q_i p_i \\
%L_1 \land L_2 \to l_1 l_2 \\
%\lnot \literal_i \equiv 
\myconjugate{\literal}_i & \to & \Identity - q_i p_i \\
\literal_i \land \literal_j & \to & q_i p_i \, q_j p_j
\end{array}
\end{equation}
%\begin{equation}
%\begin{array}{ll}
%\mathrm{F} \to 0 & \mathrm{T} \to \Identity \\
%\literal_i \to q_i p_i & \myconjugate{\literal}_i \to p_i q_i \\
%\literal_i \land \literal_j \to \literal_i \literal_j & \lnot\literal_i \to \Identity - \literal_i \qquad \mbox{that imply} \\
%\literal_i \lor \literal_j \to \literal_i + \literal_j - \literal_i \literal_j
%\end{array}
%\end{equation}
%where the rules for the substitution of the logical operations AND and NOT obviously generalize from literals to any Boolean expression.\newline
the rules for the substitutions of the logical AND and NOT holding for any Boolean expression and not just for literals. With $\anticomm{p_{i}}{q_{i}} = \Identity$ and with De Morgan's relations we easily get
\begin{equation}
\label{Boolean_substitutions_2}
\begin{array}{lll}
\myconjugate{\literal}_i & \to & p_i q_i \\
\literal_i \lor \literal_j & \to & q_i p_i + q_j p_j - q_i p_i \, q_j p_j = \literal_i + \literal_j - \literal_i \literal_j
\end{array}
\end{equation}
where we take the liberty to use freely $\literal_i$ also in $\myClg{}{}{\R^{n,n}}$ since it identifies with $q_i p_i$ with Clifford product standing for logical AND.

With a simple exercise we see that all elements of $\myClg{}{}{\R^{n,n}}$ of (\ref{Boolean_substitutions}) and (\ref{Boolean_substitutions_2}) are idempotents, including the expression of the logical OR, and thus, by proposition~\ref{S_properties}, are in $\mysetS$. Since the set $\mysetS$ is closed under multiplication also any $1\SAT$ formula is in $\mysetS$ and an idempotent. At this point it is manifest that the $2^n$ primitive idempotents of the basis $\{ \myprimidemp_i \}$ are in one to one correspondence with the possible $2^n$ $1\SAT$ formulas of the $n$ literals (Boolean \emph{atoms}), for example:
\opt{margin_notes}{\mynote{mbh.note: here could go a remark on $\lor$ going to $+$ for orthogonal expressions, \eg $\myprimidemp_i$}}%
$$
\literal_1 \myconjugate{\literal}_2 \cdots \literal_n \to q_1 p_1 \, p_2 q_2 \cdots q_n p_n \dotinformula
$$
% is one of the primitive idempotents of (\ref{formula_identity_def}).
% and one term of a generic full DNF expansion.

It is also instructive to derive that \eg $q_1 p_1 \in \mysetS$ directly from EFB formalism; with (\ref{formula_identity_def})
\begin{equation}
\label{literal_projection}
q_1 p_1 = q_1 p_1 \Identity = q_1 p_1 \prod_{j = 2}^{n} \anticomm{q_j}{p_j}
\end{equation}
since $q_1 p_1 \anticomm{q_1}{p_1} = q_1 p_1$ and the full expansion is a sum of $2^{n - 1}$ EFB terms that are all primitive idempotents and thus $q_1 p_1 \in \mysetS$ the sum being precisely the expansion in the basis $\{ \myprimidemp_i \}$. From the logical viewpoint this can be interpreted as the property that given the $1\SAT$ formula $\literal_1$ the other, unspecified, $n-1$ literals $\literal_2, \ldots, \literal_n$ are free to take all possible $2^{n - 1}$ values or, more technically, that the DNF $\literal_1$ has a \emph{full} DNF with $2^{n - 1}$ terms.

We prove now the general result that maps any Boolean expression with $n$ Boolean variables in $\myClg{}{}{\R^{n,n}}$:
\begin{MS_Proposition}
\label{logical_formulas_in_S}
Any Boolean expression $\myBooleanS$ with $n$ Boolean variables is represented in $\myClg{}{}{\R^{n,n}}$ by $S \in \mysetS$ by means of substitutions (\ref{Boolean_substitutions}) moreover $\myconjugate{\myBooleanS}$ is represented by $\Identity - S$ both being idempotents. Given another Boolean expression ${\cal Q}$ the logical equivalence $\myBooleanS \equiv {\cal Q}$ holds if and only if for their respective idempotents $S = Q$.
\end{MS_Proposition}
\begin{proof}

Any Boolean expression can be rewritten using only logical AND and NOT and thus we need only to check that (\ref{Boolean_substitutions}) reproduces these two functions faithfully in $\myClg{}{}{\R^{n,n}}$. So we verify that $\literal_i \land \literal_j$ is $\mathrm{F}$ if and only if, for the corresponding expression in $\myClg{}{}{\R^{n,n}}$, $\literal_i \literal_j = 0$.

To do so we use the property that any Boolean expression $\myBooleanS$ is $\mathrm{T}$ for a given assignment of its $n$ literals, expressed as a $1\SAT$ formula, \eg $\literal_1 \literal_2 \cdots \literal_n$, if and only if $\myBooleanS \land \literal_1 \literal_2 \cdots \literal_n$ is $\mathrm{T}$ and we may thus interpret this formula as the \emph{substitution} of $\literal_1 \literal_2 \cdots \literal_n$ into $\myBooleanS$. Put in another way considering $\myBooleanS$ in full DNF this formula selects term $\literal_1 \literal_2 \cdots \literal_n$ (if it is present) a kind of projection, a term that will turn out to be exact in $\myClg{}{}{\R^{n,n}}$.

In $\myClg{}{}{\R^{n,n}}$ for all possible substitutions for $\literal_i \literal_j$, namely $\literal_i \literal_j$, $\literal_i \myconjugate{\literal}_j$, $\myconjugate{\literal}_i \literal_j$ or $\myconjugate{\literal}_i \myconjugate{\literal}_j$, only the first one is non zero and thus the logical function AND is correctly mapped to Clifford product. Similarly for the logical NOT: to $\myconjugate{\literal}_i$ corresponds $\Identity - \literal_i$ and by idempotent properties $(\Identity - \literal_i) \literal_i = 0$ and, interpreting the first term as the expression of $\myconjugate{\literal}_i$ in $\myClg{}{}{\R^{n,n}}$ and the second as a substitution of Boolean variables, the proof is complete when we observe that $\literal_i$ is the only of the two possible assignments giving this result.

Starting from Boolean expression $\myBooleanS$ written only with AND and NOT we see that the corresponding $S \in \myClg{}{}{\R^{n,n}}$ obtained by (\ref{Boolean_substitutions}) is in $\mysetS$ since we already noticed that all literals are in $\mysetS$ and that the set is closed under multiplication and under inner operation $s \to \Identity - s$.

For the part on logical equivalence $\myBooleanS \equiv {\cal Q}$ we consider the expressions in their full DNF and we just remarked that $\myBooleanS \land \literal_1 \literal_2 \cdots \literal_n$ selects the corresponding DNF term if it exists, thus logical equivalence holds if and only if they are equal term by term. With rules (\ref{Boolean_substitutions}) we obtain two expressions $S, Q \in \mysetS$ that have unique expansions (\ref{formula_S_def}) in the basis of primitive idempotents $\{ \myprimidemp_i \}$. By standard properties of the basis of a linear space, $S = Q$ is equivalent to the equality of their base expansions that can be checked component by component by $\myprimidemp_i S = \myprimidemp_i Q$ that is clearly equivalent to the termwise check of the full DNF of $\myBooleanS$ and ${\cal Q}$.
\opt{margin_notes}{\mynote{mbh.note: this proves the unicity of the full DNF}}%
\end{proof}

We remark that the full equivalence between Boolean expressions and elements of $\myClg{}{}{\R^{n,n}}$ holds \emph{only} for elements of $\mysetS$ that is not closed under addition so we will have to use algebra with some care. For example $S_1 + S_2 = Q$ has a logical meaning if all terms belong to $\mysetS$ and in this case this relation is equivalent to \eg $S_1 = Q - S_2$ to which we may attach a logical meaning, on the other hand $S_1 + 2 S_2 \notin \mysetS$ and strictly has no logical counterpart.

A useful property is that, for every $1 \le i \le n$, $\myAbelsa$, as a vector space, may be written as a direct sum of two orthogonal subspaces (and subalgebras)
\begin{equation}
\label{formula_P_subspaces}
\myAbelsa = \myAbelsa_i \oplus \myconjugate{\myAbelsa}_i = \literal_i \myAbelsa \oplus \myconjugate{\literal}_i \myAbelsa
% q_i p_i \myAbelsa \oplus p_i q_i \myAbelsa
\end{equation}
\opt{margin_notes}{\mynote{mbh.ref: for a proof see log p. 695}}%
where $\myAbelsa_i$ and $\myconjugate{\myAbelsa}_i$ are subspaces of dimension $2^{n - 1}$ obtained by projections $\myAbelsa_i = q_i p_i \myAbelsa = \literal_i \myAbelsa$ as is simple to verify writing any element of $\myAbelsa$ in $\{ \myprimidemp_i \}$ basis and remembering that $q_i p_i \anticomm{q_i}{p_i} = q_i p_i$, a projection already applied in (\ref{literal_projection}). Both subspaces contain corresponding orthogonal subsets $\mysetS_i$ and $\myconjugate{\mysetS}_i$ and, by (\ref{formula_S_def}), for any $s \in \mysetS_i$ then $\Identity - s \in \myconjugate{\mysetS}_i$.

\section{Satisifability in Clifford algebra}
\label{SAT_reformulation}
Previous results apply to any Boolean expression and thus to $\SAT$ formulas but we give a more immediate way of expressing \SAT{} problems in $\myClg{}{}{\R^{n,n}}$ tailored to our needs:
\begin{MS_Proposition}
\label{SAT_in_Cl_2}
Given a SAT problem $\myBooleanS$ (\ref{formula_SAT_std}) with $m$ clauses ${\cal C}_j \equiv (\literal_{j_1} \lor \literal_{j_2} \lor \cdots \lor \literal_{j_k})$, for each clause let $z_j := \myconjugate{\literal}_{j_1} \myconjugate{\literal}_{j_2} \cdots \myconjugate{\literal}_{j_k}$, then $\myBooleanS$ is satisfiable if and only if, the corresponding algebraic expression of $\myClg{}{}{\R^{n,n}}$
\begin{equation}
\label{formula_SAT_EFB_2}
S = \prod_{j = 1}^m (\Identity - z_j) \ne 0 \dotinformula
\end{equation}
\end{MS_Proposition}

\begin{proof}
With De Morgan's relations and (\ref{Boolean_substitutions}) we get that clause ${\cal C}_j$ is represented in $\myClg{}{}{\R^{n,n}}$ by
$$
{\cal C}_j \to \Identity - \myconjugate{\literal}_{j_1} \myconjugate{\literal}_{j_2} \cdots \myconjugate{\literal}_{j_k} = \Identity - z_j
$$
and by previous results ${\cal C}_j$ is $\mathrm{F}$ for all and only the assignments giving $\Identity - z_j = 0$ in $\myClg{}{}{\R^{n,n}}$.

By the substitution mechanism mentioned in the proof of proposition~\ref{logical_formulas_in_S} for all and only the assignments $\literal_1 \literal_2 \cdots \literal_n$ that satisfy $\myBooleanS$ then $\myBooleanS \land \literal_1 \literal_2 \cdots \literal_n$ is $\mathrm{T}$ and thus, necessarily, ${\cal C}_j \land \literal_1 \literal_2 \cdots \literal_n$ is $\mathrm{T}$ for all $j$. In $\myClg{}{}{\R^{n,n}}$ this is equivalent to $(\Identity - z_j) \literal_1 \literal_2 \cdots \literal_n = \literal_1 \literal_2 \cdots \literal_n \ne 0$ for all $j$ and thus also $S \literal_1 \literal_2 \cdots \literal_n \ne 0$.
%It is clear that for any $j$ $z_j \in \mysetS$ and thus also $\Identity - z_j \in \mysetS$ and thus also $S \in \mysetS$.
\end{proof}

We remark that expanding the product of two clauses of (\ref{formula_SAT_EFB_2}) we get
\begin{equation}
\label{formula_z_j_expansion}
(\Identity - z_j) (\Identity - z_l) = \Identity - z_j - z_l + z_j z_l
\end{equation}
and $z_j z_l = 0$ if and only if in $z_j$ and $z_l$ appears the same literal in opposite forms ($\literal_i \myconjugate{\literal}_i = 0$). In any way this product is always in $\mysetS$ even if the generic terms of the expansion are not in general in $\mysetS$, \eg $- z_j$. In general the product of $m$ clauses will expand to at most $2^m$ terms that is nothing else than the Boolean expansion of $\myBooleanS$ in its DNF of section~\ref{SAT_basics} \emph{and} the expansion of $S$ in basis $\{ \myprimidemp_i \}$. The first term of the expansion is certainly $\Identity$ (\ref{formula_z_j_expansion}) so that, calling $\Delta$ the other terms we rewrite (\ref{formula_SAT_EFB_2}) as
\begin{equation}
\label{formula_SAT_EFB_3}
S = \prod_{j = 1}^m (\Identity - z_j) := \Identity - \Delta
\end{equation}
and since $S \in \mysetS$ also $\Delta = \Identity - S$ is in $\mysetS$ and all elements of this relation are idempotents.

\section{A condition for unsatisfiability}
\label{unSAT_condition}
Exploiting this \SAT{} formulation we prove the main result of this paper:
\opt{margin_notes}{\mynote{mbh.note: probably the result is correct also for \SAT{} in $\myAbelsa$ with OR mapped to $+$}}%
\begin{MS_theorem}
\label{SAT_unsat_thm}
A given nonempty \SAT{} problem in $\myClg{}{}{\R^{n,n}}$ (\ref{formula_SAT_EFB_2}) admits no solution if and only if, for all generators (\ref{formula_generators}) of $\myClg{}{}{\R^{n,n}}$
\begin{equation}
\label{SAT_symmetry_formula}
\mygen_i \; S \; \mygen_i^{-1} = S \qquad \forall \; 1 \le i \le 2 n \dotinformula
\end{equation}
\end{MS_theorem}
\begin{proof}
We know that $S \in \mysetS$; a simple consequence of proposition~\ref{literal_invariance} in next section is that also $\mygen_i \; S \; \mygen_i^{-1} \in \mysetS$ so both terms are in $\mysetS$.
%, by proposition~\ref{logical_formulas_in_S}, it is enough to verify the algebraic equality of the expressions.

By (\ref{formula_SAT_EFB_3}) $\mygen_i \; S \; \mygen_i^{-1} = S$ if and only if $\mygen_i \Delta \mygen_i^{-1} = \Delta$ and by a fundamental property of Clifford algebra \cite[Propostion~16.6]{Porteous_1995} an element of the%
\opt{margin_notes}{\mynote{mbh.note: probably one could also say more simply that $\myClg{}{}{\R^{n,n}}$ is central}}%
{} even subalgebra commutes with all generators $\mygen_i$ if and only it is of the form $\delta \Identity$ for $\delta \in \R$. So $\mygen_i \Delta \mygen_i^{-1} = \Delta$ for all $i$ is equivalent to $\Delta = \delta \Identity$ but we know also that $\Delta \in \mysetS$ and thus necessarily $\delta \in \{0, 1\}$. By (\ref{formula_SAT_EFB_3}) $\delta = 0$ would imply $S = \Identity$ that represents a \SAT{} problem with no clauses, excluded by hypothesis, so $\Delta = \Identity$ that implies $S = 0$ that is thus unsatisfiable.
%
%
%
%
%
%\noindent old demo
%
%
%
%By proposition~\ref{SAT_in_Cl_2} we know that \SAT{} admits no solution only if $\prod_{j = 1}^m (\Identity - z_j) = 0$ and by (\ref{formula_SAT_EFB_3}) this is equivalent to $\Delta = \Identity$. It is a standard property of Clifford algebra \cite[Propostion~16.6]{Porteous_1995} that an element of the algebra can be written as $\delta \Identity$ for $\delta \in \R$ if and only if it commutes with all generators $\mygen_i$. Assuming $\mygen_i \Delta \mygen_i^{-1} = \Delta$ we need to prove that $\delta = 1$ but we already know that $\Delta \in \mysetS$ and if also $\Delta = \delta \Identity$ necessarily $\delta = 1$; obviously the condition $\mygen_i \Delta \mygen_i^{-1} = \Delta$ is equivalent to the thesis.
%
%
%\noindent old old demo
%
%
%
%With (\ref{formula_SAT_EFB_3}) clearly this is equivalent to prove that $\mygen_i \Delta \mygen_i^{-1} = \Delta$ and applying proposition~\ref{SAT_in_Cl_2} to (\ref{formula_SAT_EFB_3}) we see that the problem is unsatisfiable if and only if $\Delta = \Identity$. It is a standard property of Clifford algebra \cite[Propostion~16.6]{Porteous_1995} that an element of the algebra commutes with all generators $\mygen_i$ if and only if it may be written as $\delta \Identity$ for $\delta \in \R$. To complete the proof we need to prove that $\delta = 1$ but we already know that $\Delta \in \mysetS$ and if also $\Delta = \delta \Identity$ necessarily $\delta = 1$.
\end{proof}

It is known that a given \SAT{} problem $\myBooleanS$ is unsatisfiable if and only if $\myconjugate{\myBooleanS}$ is a tautology, namely if $\myconjugate{\myBooleanS} \equiv \mathrm{T}$. By proposition~\ref{logical_formulas_in_S} and (\ref{formula_SAT_EFB_3}) $\myconjugate{S} = \Identity - S = \Delta$ that gives the algebraic version of this property in $\mysetS$ and we have
\begin{MS_Corollary}
\label{TAUT_in_Cl}
A Boolean expression $\myconjugate{\myBooleanS}$ is a tautology if and only if for its corresponding idempotent $\Delta \in \mysetS$, $\Delta \ne 0$ and
$$
\mygen_i \; \Delta \; \mygen_i^{-1} = \Delta \qquad \forall \; 1 \le i \le 2 n
$$
\end{MS_Corollary}
\noindent and thus $\Identity \in \mysetS$ represent a tautology and (\ref{formula_identity_def}) is its full DNF.

%With (\ref{formula_SAT_EFB_3}) the condition of the theorem~\ref{SAT_unsat_thm} may be written as
%\begin{equation}
%\label{formula_invariance_condition_2}
%\mygen_i \; S \; \mygen_i^{-1} = \prod_{j = 1}^m (\Identity - \mygen_i z_j \mygen_i^{-1}) = \prod_{j = 1}^m (\Identity - z_j) \qquad 1 \le i \le 2 n \dotinformula
%\end{equation}
%\noindent We remark that in general this does \emph{not mean} that $\mygen_i z_j \mygen_i^{-1} = z_j$ but this form will result nevertheless useful. To apply this result to actual \SAT{} literals it is easier to use the isomorphic $\R(2^n)$ matrix algebra.

We remark that in $\myClg{}{}{\R^{n,n}}$ tautologies and unsatisfiable \SAT{}, corresponding respectively to $S = \Identity, 0$, are the only instances of \SAT{} that have the maximally symmetric form of the scalar elements of $\myClg{}{}{\R^{n,n}}$ (\ref{SAT_symmetry_formula}).

\section{An algorithm to test for unsatisfiability}
\label{unSAT_algorithm}
We exploit results of previous section to outline a \SAT{} algorithm: its basic idea is that if a given \SAT{} problem is satisfiable it infringes at least one of the $2 n$ symmetry relations of theorem~\ref{SAT_unsat_thm}. We begin with a simple result:

\begin{MS_Proposition}
\label{literal_invariance}
For any literal $\literal_i$ in $\myClg{}{}{\R^{n,n}}$
$$
\begin{array}{l}
\mygen_j \; \literal_i \; \mygen_j^{-1} = \myconjugate{\literal}_i \qquad \mbox{for} \qquad j \in \{2 i - 1, 2i\} \\
\mygen_j \; \literal_i \; \mygen_j^{-1} = \literal_i \qquad \mbox{for} \qquad j \notin \{2 i - 1, 2i\}
\end{array}
$$
\end{MS_Proposition}
\noindent proved using (\ref{formula_generators}) and (\ref{formula_Witt_basis}). Thus, when applying theorem~\ref{SAT_unsat_thm}, it is sufficient to test invariance just for the odd generators $\mygen_{2 i - 1}$ for which $\mygen_{2 i - 1}^{-1} = \mygen_{2 i - 1}$ and the literal $\literal_i$ is complemented by ``its'' generator $\mygen_{2 i - 1}$ and left invariant by all others, a typical case of reflections in Clifford algebras.

With this result we can check theorem~\ref{SAT_unsat_thm} in a simple example with $n = 1$: the only unsatisfiable \SAT{}, $\literal_1 \myconjugate{\literal}_1$, is actually invariant since $\mygen_1 \literal_1 \myconjugate{\literal}_1 \mygen_1 = \literal_1 \myconjugate{\literal}_1$. The theorem now looks sensible also in the Boolean formulation of \SAT{}: a problem is unsatisfiable if and only if exchanging any literal for its complement the result is not altered.

To derive an actual algorithm from these relations we need a technical
\begin{MS_lemma}
\label{SAT_recursive_expansion}
Given a non empty $k\SAT$ problem $S$ with $n$ literals, for any literal $\literal_i$ we may write $S$ as
\begin{equation}
\label{formula_SAT_recursive_expansion}
S = \myconjugate{\literal}_i S_{\underline{i} 1} + \literal_i S_{\underline{i} 2}
\end{equation}
\opt{margin_notes}{\mynote{mbh.note: here $+$ coincides with the logical OR since terms belong to orthogonal subspaces (old proof)}}%
where $S_{\underline{i} 1}$ and $S_{\underline{i} 2}$ are $\SAT$ problems, represented in $\mysetS$, with $n - 1$ literals: all those of $S$ but the $i$-th, this being the reason for notation $\underline{i}$. More in detail:
$$
S_{\underline{i} 1} = S_{\underline{i} 0} S_{\underline{i} 1}' \qquad S_{\underline{i} 2} = S_{\underline{i} 0} S_{\underline{i} 2}'
$$
where:
\begin{itemize}
\item[-] $S_{\underline{i} 0}$ is the $k\SAT$ problem formed by the subset of clauses of $S$ that do not contain neither $\literal_i$ nor $\myconjugate{\literal}_i$;
\item[-] $S_{\underline{i} 1}'$ is the $(k-1)\SAT$ problem formed by the subset of clauses of $S$ that contained $\myconjugate{\literal}_i$ and in which $\myconjugate{\literal}_i$ have been removed;
\item[-] $S_{\underline{i} 2}'$ is the $(k-1)\SAT$ problem formed by the subset of clauses of $S$ that contained $\literal_i$ and in which $\literal_i$ have been removed.
\end{itemize}
\end{MS_lemma}

\begin{proof}
We have seen that for any literal $\literal_i$ we can decompose $\myAbelsa$ in a direct sum of orthogonal subspaces (\ref{formula_P_subspaces}), it follows that any $S \in \mysetS$ has a unique decomposition with components belonging to the two subspaces; the components are obtained by projections $\myconjugate{\literal}_i S$ and $\literal_i S$ so that $S = \myconjugate{\literal}_i S + \literal_i S$, moreover both terms are in $\mysetS$. Calculating explicitly the terms of (\ref{formula_SAT_recursive_expansion}) with the projections is immediate with $S$ in the form (\ref{formula_SAT_EFB_2}).
% or also, with the substitution mechanism mentioned in the proof of proposition~\ref{logical_formulas_in_S}, substituting $\literal_i \equiv \mathrm{T}, \mathrm{F}$ in $S$.
\end{proof}

\begin{MS_Proposition}
\label{e_i_invariance}
When testing the invariance of a given Boolean expression $S$ for any single generator $\mygen_{2 i - 1}$ for $1 \le i \le n$ the three following statements are equivalent
\begin{equation}
\label{formula_e_i_invariance}
\begin{array}{r l l}
\mygen_{2 i - 1} \; S \; \mygen_{2 i - 1} & = & S \\
S_{\underline{i} 1} & = & S_{\underline{i} 2} \\
S & = & (\literal_i + \myconjugate{\literal}_i) S_{\underline{i} 1} = S_{\underline{i} 1}
\end{array}
\end{equation}
with $S_{\underline{i} 1}$ and $S_{\underline{i} 2}$ defined as in lemma~\ref{SAT_recursive_expansion}.
\end{MS_Proposition}

\begin{proof}
To prove the proposition we prove that any statement implies the successive, circularly. Along the proof we will see that all terms of these relations are in $\mysetS$ and so the use of algebraic equality is consistent.

Assuming the first statement and writing $S$ in form (\ref{formula_SAT_recursive_expansion}) we get
$$
\mygen_{2 i - 1} (\myconjugate{\literal}_i S_{\underline{i} 1} + \literal_i S_{\underline{i} 2}) \mygen_{2 i - 1} = \myconjugate{\literal}_i S_{\underline{i} 2} + \literal_i S_{\underline{i} 1}
$$
since $S_{\underline{i} 1}$ and $S_{\underline{i} 2}$ are without the $i$-th literal and thus, by proposition~\ref{literal_invariance}, are left invariant by $\mygen_{2 i - 1}$.
%
%
%clearly for problems $S_{\underline{i} 0}, S_{\underline{i} 1}$ and $S_{\underline{i} 2}$ for all their clauses in form (\ref{formula_SAT_EFB_3}) by proposition~\ref{literal_invariance} $\mygen_{2 i - 1} z_j \mygen_{2 i - 1} = z_j$ so they are all invariant and thus out of the game when testing $\mygen_{2 i - 1}$ invariance. So it's easy to conclude that
%$$
%\mygen_{2 i - 1} S \mygen_{2 i - 1} = \myconjugate{\literal}_i S_{\underline{i} 0} S_{\underline{i} 2} + \literal_i S_{\underline{i} 0} S_{\underline{i} 1} \stackrel{?}{=} \myconjugate{\literal}_i S_{\underline{i} 0} S_{\underline{i} 1} + \literal_i S_{\underline{i} 0} S_{\underline{i} 2}
%$$
%where the second equality is our hypothesis.
The two terms belong to the two orthogonal subspaces $\literal_i \myAbelsa$ and $\myconjugate{\literal}_i \myAbelsa$ of $\myAbelsa$ (\ref{formula_P_subspaces}). It follows that equality with $\myconjugate{\literal}_i S_{\underline{i} 1} + \literal_i S_{\underline{i} 2}$ holds only if it holds separately in the two subspaces and this happens iff
$$
S_{\underline{i} 1} = S_{\underline{i} 2} \dotinformula
$$
Assuming this relation with $S$ in form (\ref{formula_SAT_recursive_expansion}) we immediately get with (\ref{Boolean_substitutions})
$$
S = (\literal_i + \myconjugate{\literal}_i) S_{\underline{i} 1} = S_{\underline{i} 1}
$$
and in turn this form guarantees $\mygen_{2 i - 1} \; S \; \mygen_{2 i - 1} = S$.
\end{proof}

With these results we outline a recursive algorithm for $k\SAT$ problem $S$: we select a literal $\literal_i$, calculate $S$ in form (\ref{formula_SAT_recursive_expansion}) and test wether $S_{\underline{i} 1} = S_{\underline{i} 2}$%
\footnote{we remark that using the more detailed version of (\ref{formula_SAT_recursive_expansion}) $S = \myconjugate{\literal}_i S_{\underline{i} 0} S_{\underline{i} 1}' + \literal_i S_{\underline{i} 0} S_{\underline{i} 2}'$ if $S_{\underline{i} 0}$ is invertible condition $S_{\underline{i} 0} S_{\underline{i} 1}' = S_{\underline{i} 0} S_{\underline{i} 2}'$ is equivalent to $S_{\underline{i} 1}' = S_{\underline{i} 2}'$ that is usually a much simpler problem since $S_{\underline{i} 1}'$ and $S_{\underline{i} 2}'$ are $(k-1)\SAT$ problems. But $S_{\underline{i} 0} \in \mysetS$ and thus it is invertible only if $S_{\underline{i} 0} = \Identity$, namely if $S_{\underline{i} 0}$ has no clauses. Nevertheless the condition $S_{\underline{i} 1}' = S_{\underline{i} 2}'$ is sufficient to get $S_{\underline{i} 0} S_{\underline{i} 1}' = S_{\underline{i} 0} S_{\underline{i} 2}'$ but in general not necessary.}%
. If the answer is no by theorem~\ref{SAT_unsat_thm} and proposition~\ref{e_i_invariance} we have a certificate that $S$ is satisfiable. If the answer is yes by proposition~\ref{e_i_invariance} we can write $S = S_{\underline{i} 1}$ and we restart with this new problem in $n-1$ literals. If the process continues for all the $n$ literals we fulfill the conditions of theorem~\ref{SAT_unsat_thm} proving that $S$ has the fully symmetric form of the scalars (\ref{formula_identity_def}) in $\myClg{}{}{\R^{n,n}}$ and is thus unsatisfiable.

In practice it can be difficult to prove or disprove that $S_{\underline{i} 1} = S_{\underline{i} 2}$ and we need to use a slightly more complex strategy based on the following principle: $S_{\underline{i} 1} = S_{\underline{i} 2}$ is equivalent to $S_{\underline{i} 1} - S_{\underline{i} 2} = 0$ and since this is an algebraic expression in $n-1$ literals we can check that it is a fully symmetric expression, namely a scalar in Clifford algebra $\myClg{}{}{\R^{n-1,n-1}}$.

%\bigskip

We conclude showing that this algorithm constitutes a \emph{relaxation} of the Davis Putnam recursive $\SAT$ algorithm (DPLL) based on results we resume here, in our notation to ease comparisons \cite[Theorem~4.1 p.~243]{Davis_Weyuker_1983}:
\begin{enumerate}
\item any $k\SAT$ problem $\myBooleanS$ can be written, with respect to any of its literals $\literal_i$, in the form (lemma~\ref{SAT_recursive_expansion} gives an alternative algebraic proof of this)
$$
\myBooleanS \equiv \myconjugate{\literal}_i \myBooleanS_{\underline{i} 1} \lor \literal_i \myBooleanS_{\underline{i} 2}
$$
\item $\myBooleanS$ is unsatisfiable if and only if for all literals $\literal_i$ the derived problems $\myBooleanS_{\underline{i} 1}$ and $\myBooleanS_{\underline{i} 2}$ are both unsatisfiable $\myBooleanS_{\underline{i} 1} \equiv \myBooleanS_{\underline{i} 2} \equiv \mathrm{F}$, in our notation
\begin{equation}
\label{formula_sat_recursion}
S_{\underline{i} 1} = S_{\underline{i} 2} = 0 \dotinformula
\end{equation}
\end{enumerate}

These properties are at the heart of a recursive algorithm very similar to our one with the noticeable difference that condition (\ref{formula_sat_recursion}) is stronger than (\ref{formula_e_i_invariance}) and in this sense our algorithm is a relaxation of DPLL.
\opt{margin_notes}{\mynote{mbh.note: and provides a geometrical interpretation of DPLL}}%

In other words $S_{\underline{i} 1} = S_{\underline{i} 2} \ne 0$ does not satisfy (\ref{formula_sat_recursion}) but satisfies (\ref{formula_e_i_invariance}) and we can proceed to test invariance with respect to the next literal. Only if the procedure continues up to the last literal, by theorem~\ref{SAT_unsat_thm}, we will have proved that $S$ is a scalar and thus $S = 0$ and so only if we arrive at the end we will have proved that at every step we actually had not only $S_{\underline{i} 1} = S_{\underline{i} 2}$ but $S_{\underline{i} 1} = S_{\underline{i} 2} = 0$. A more detailed version of this algorithm together with a formal proof of its correctness is contained in a dedicated appendix.

%A final remark is that lemma~\ref{SAT_recursive_expansion} provides a purely algebraic proof of Davis Putnam pivotal result \cite[Theorem~4.1]{Davis_Weyuker_1983}.

\section{Conclusions}
\label{Conclusions}
It is not clear wether these results can produce an algorithm competitive with state of the art \SAT{} solvers but its theoretical properties look interesting at least for the transformation of a typical combinatorial problem, it is tempting to say \emph{the} archetypical combinatorial problem, in a purely algebraic setting in Clifford algebra.

The main contribution of Clifford algebra is that it allows to deal with the exponentially large DNF expressions of \SAT{} problems algebraically. Moreover testing unsatisfiability by means of geometric invariance of idempotents under reflections of $\mygen_{2 i - 1}$ generators appears promising.

\SAT{} formulation (\ref{formula_SAT_EFB_3}) can be exploited to find also other conditions guaranteeing unsatisfiability, namely $\myconjugate{S} = \Delta = \Identity$. For example in the isomorphic matrix algebra $\R(2^n)$, $\Delta$ can be always a diagonal matrix and thus the problem is satisfiable if and only if $\det \left( \Delta \right) = 0$. With Clifford algebra we can manipulate these matrices of size $2^n$ manipulating just $n$ matrices of size 2, \eg with Jordan-Wigner construction.

We conclude noting that Clifford algebra, beyond its ability to represent automorphisms of linear spaces, is well suited also to deal with Boolean expressions.

%\newpage

\section*{Appendix}
In section~\ref{unSAT_algorithm} we outlined a \SAT{} algorithm that exploits the relaxation of (\ref{formula_sat_recursion}) to (\ref{formula_e_i_invariance}): we describe it here together with a proof of its correctness.

We start proving two propositions that extend theorem~\ref{SAT_unsat_thm}: the first one to milder conditions and the second one to generalized variables.
\begin{MS_Proposition}
\label{SAT_recursion_algorithm}
A non empty $\SAT$ problem $S$ is unsatisfiable if and only if for all literals $\literal_i$ and their relative expansion (\ref{formula_SAT_recursive_expansion})
\begin{equation}
\label{formula_sat_necessary_condition}
S_{\underline{i} 1} - S_{\underline{i} 2} = \delta \Identity \qquad \delta \in \R \qquad \forall \; 1 \le i \le n \dotinformula
\end{equation}
\end{MS_Proposition}
\begin{proof}
Assuming $S = 0$ by theorem~\ref{SAT_unsat_thm} and proposition~\ref{e_i_invariance} for any literal $\literal_i$ we have $S_{\underline{i} 1} - S_{\underline{i} 2} = 0$. To prove the converse let $S_{\underline{i} 1} - S_{\underline{i} 2} = \delta \Identity$ with $\delta \in \R$; since both $S_{\underline{i} 1}$ and $S_{\underline{i} 2}$ are in $\mysetS$ and belong to a $2^{n - 1}$-dimensional subspace of $\myAbelsa$ (\ref{formula_P_subspaces}) we easily see that equality can hold only for $\delta = 0$, thus $S_{\underline{i} 1} = S_{\underline{i} 2}$ and since this holds for all literals $\literal_i$, by theorem~\ref{SAT_unsat_thm}, $S = 0$.
\opt{margin_notes}{\mynote{mbh.note: here commented there is an old, longer and possibly better proof, indeed $S_{\underline{i} j}$ are in $\myClg{}{}{\R^{n-1,n-1}}$}}%
\end{proof}
%
%\begin{proof}
%%Assuming $S = 0$ by theorem~\ref{SAT_unsat_thm} and proposition~\ref{e_i_invariance} for any literal $\literal_i$ we have $S_{\underline{i} 0} S_{\underline{i} 1} - S_{\underline{i} 0} S_{\underline{i} 2} = 0$.
%To prove the converse let us assume that $S_{\underline{i} 0} S_{\underline{i} 1} - S_{\underline{i} 0} S_{\underline{i} 2} = \delta \Identity$ with $\delta \in \R$; since $S$ is non empty at least one of the two terms $S_{\underline{i} 0} S_{\underline{i} 1}$ and $S_{\underline{i} 0} S_{\underline{i} 2}$ must not be a tautology, let it be $S_{\underline{i} 0} S_{\underline{i} 1}$ and let us write
%$$
%S_{\underline{i} 0} S_{\underline{i} 1} = \delta \Identity + S_{\underline{i} 0} S_{\underline{i} 2}
%$$
%since both $S_{\underline{i} 0} S_{\underline{i} 1}$ and $S_{\underline{i} 0} S_{\underline{i} 2}$ are in $\mysetS$ we conclude that necessarily $\delta \in \{0, 1\}$. Supposing $\delta = 1$ since $\Identity + S_{\underline{i} 0} S_{\underline{i} 2} \in \mysetS$ necessarily $S_{\underline{i} 0} S_{\underline{i} 2} = 0$ that would imply $S_{\underline{i} 0} S_{\underline{i} 1} = \Identity$ but this is impossible since we supposed that $S_{\underline{i} 0} S_{\underline{i} 1}$ is not a tautology. We must conclude that the only possibility is $\delta = 0$ namely $S_{\underline{i} 0} S_{\underline{i} 1} = S_{\underline{i} 0} S_{\underline{i} 2}$ and since this must hold for any literal $\literal_i$ it follows, by quoted propositions, that $S = 0$.
%\end{proof}

To deal with expressions like $S_{\underline{i} 1} - S_{\underline{i} 2}$ in full generality we introduce a new kind of variables $\myGenVarS$ of the form
\begin{equation}
\label{formula_signed_sum}
\myGenVarS = \sum_{j = 1}^{r} s_j S_{j} \qquad s_j = \pm 1, \; S_{j} \in \mysetS
\end{equation}
where $S_{j}$ are \SAT{} problems with $n$ literals. In general $S_{j} \in \mysetS$ and $\myGenVarS \in \myAbelsa$, being a sum of signed idempotents, but in general $\myGenVarS \notin \mysetS$. Also to $\myGenVarS$ we may apply decomposition (\ref{formula_SAT_recursive_expansion}), applying it first to the \SAT{} problems of the sum $S_{j}$ getting, with straightforward extension of notation,
\begin{equation}
\label{formula_decomposition_signed_sum}
\myGenVarS = \sum_{j = 1}^{r} s_j \left(\myconjugate{\literal}_i S_{j \underline{i} 1} + \literal_i S_{j \underline{i} 2}\right) := \myconjugate{\literal}_i \myGenVarS_{\underline{i} 1} + \literal_i \myGenVarS_{\underline{i} 2}
\end{equation}
where we defined two new variables of the new kind, in $n-1$ literals
$$
\myGenVarS_{\underline{i} 1} = \sum_{j = 1}^{r} s_j S_{j \underline{i} 1} \qquad \myGenVarS_{\underline{i} 2} = \sum_{j = 1}^{r} s_j S_{j \underline{i} 2} \dotinformula
$$
Now we extend theorem~\ref{SAT_unsat_thm} to the new variables $\myGenVarS$ proving the slightly weaker relation $\myGenVarS = \delta \Identity$, with $\delta \in \R$.
\begin{MS_Proposition}
\label{SAT_list_proposition}
Give a signed sum $\myGenVarS$ of $r$ \SAT{} problems in $n$ literals (\ref{formula_signed_sum}) then $\myGenVarS = \delta \Identity$ with $\delta \in \R$ if and only if, for all its literals $\literal_i$ we have in (\ref{formula_decomposition_signed_sum})
$$
\myGenVarS_{\underline{i} 1} - \myGenVarS_{\underline{i} 2} = 0 \qquad \forall \; 1 \le i \le n
$$
where each expression $\myGenVarS_{\underline{i} 1} - \myGenVarS_{\underline{i} 2}$ has at most $2 r$ terms in $n-1$ literals.
\end{MS_Proposition}

\begin{proof}
For $r = 1$ previous results prove the thesis; for $r > 1$ we remark that theorem~\ref{SAT_unsat_thm}, being a general property of Clifford algebra, applies to any element of $\myAbelsa$ and thus to $\myGenVarS$ and thus we can proceed with algebraic equality.

Given $\myGenVarS$ we select a literal $\literal_i$ and apply the decomposition (\ref{formula_decomposition_signed_sum}) and with proposition~\ref{literal_invariance} we get
$$
\mygen_{2 i - 1} \myGenVarS \mygen_{2 i - 1} = \myconjugate{\literal}_i \myGenVarS_{\underline{i} 2} + \literal_i \myGenVarS_{\underline{i} 1}
$$
to be compared with (\ref{formula_decomposition_signed_sum}); since we are in $\myAbelsa$ by (\ref{formula_P_subspaces}) the comparison breaks up into two parts in each subspace that have to be satisfied separately and this happens if and only if $\myGenVarS_{\underline{i} 1} = \myGenVarS_{\underline{i} 2}$.
\end{proof}

\bigskip

With these two results we proceed to build a recursive algorithm; we start building a chain of implications deriving from the hypothesis that a given nonempty $\SAT$ problem $S$ is unsatisfiable, namely $S = 0$: given any literal $\literal_i$, we have the necessary condition (\ref{formula_sat_necessary_condition}), namely
\begin{equation}
\label{formula_implication_chain_1}
S = 0 \quad \stackrel{\literal_i}{\Longrightarrow} \quad S_{\underline{i} 1} - S_{\underline{i} 2} = \delta \Identity
\end{equation}
where $\literal_i$ indicates that to go from the $n$ literals of $S$ to the $n-1$ of $S_{\underline{i} 1} - S_{\underline{i} 2}$ we eliminated the $i$-th literal. This is just an implication since to prove that $S = 0$ we would need to prove the other $n-1$ similar implications, on the other hand $S_{\underline{i} 1} - S_{\underline{i} 2} \ne \delta \Identity$ proves that $S \ne 0$. The same argument can be applied to the signed sum of \SAT{} problems (\ref{formula_signed_sum}) in $\myAbelsa$:
\begin{equation}
\label{formula_implication_chain_2}
\myGenVarS = \delta \Identity \quad \stackrel{\literal_j}{\Longrightarrow} \quad \myGenVarS_{\underline{j} 1} - \myGenVarS_{\underline{j} 2} = 0 \quad \Longrightarrow \quad \myGenVarS_{\underline{j} 1} - \myGenVarS_{\underline{j} 2} = \delta \Identity \dotinformula
\end{equation}

We can connect (\ref{formula_implication_chain_1}) and (\ref{formula_implication_chain_2}) and starting from a single \SAT{} problem $S$ in $n$ literals we can build a complete chain of implications
\begin{equation}
\label{formula_implication_list}
S = 0 \;\;\; \stackrel{\literal_1}{\Longrightarrow} \;\;\; \myGenVarS_{\underline{1} 1} - \myGenVarS_{\underline{1} 2} = \delta \Identity \;\;\; \stackrel{\literal_2}{\Longrightarrow} \;\;\; \myGenVarS_{\underline{2} 1} - \myGenVarS_{\underline{2} 2} = \delta \Identity \;\;\; \stackrel{\literal_3}{\Longrightarrow} \cdots \stackrel{\literal_n}{\Longrightarrow} \;\;\; \myGenVarS_{\underline{n} 1} = \myGenVarS_{\underline{n} 2}
\end{equation}
where in the last step $\literal_n$ represent the last literal to be eliminated and thus the rightmost expression of the chain has no literals and can be easily checked for equality since at the last stage the problems can be only of two kinds: unsatisfiable problems, namely problems with empty clauses, or tautologies, namely problems with no clauses and so the final comparison $\myGenVarS_{\underline{n} 1} = \myGenVarS_{\underline{n} 2}$ can be done immediately. Obviously the choice of the literal at each step is free.

We conclude proving that the satisfaction of the $n-1$ rightmost tests of (\ref{formula_implication_list}), namely,
\begin{equation}
\label{formula_implication_list_2}
S \;\;\; \stackrel{\literal_1}{\Longrightarrow} \;\;\; \myGenVarS_{\underline{1} 1} - \myGenVarS_{\underline{1} 2} = \delta \Identity \;\;\; \stackrel{\literal_2}{\Longrightarrow} \;\;\; \myGenVarS_{\underline{2} 1} - \myGenVarS_{\underline{2} 2} = \delta \Identity \;\;\; \stackrel{\literal_3}{\Longrightarrow} \cdots \stackrel{\literal_n}{\Longrightarrow} \;\;\; \myGenVarS_{\underline{n} 1} = \myGenVarS_{\underline{n} 2}
\end{equation}
implies in turn $S = 0$ transforming the chain of implications (\ref{formula_implication_list}) in a closed circle.
\begin{MS_Proposition}
\label{implication_list_closure}
Given a non empty \SAT{} problem $S$ the implication list (\ref{formula_implication_list_2}) is fully satisfied if and only if $S = 0$.
\end{MS_Proposition}
\begin{proof}
The forward implication chain (\ref{formula_implication_list}) have already been proved. To prove the converse we prove first that if any $\myGenVarS_{\underline{i} 1} - \myGenVarS_{\underline{i} 2} \ne \delta \Identity$, namely if the chain is broken, this implies that $S$ is satisfiable: this is immediate since $\myGenVarS_{\underline{i} 1} - \myGenVarS_{\underline{i} 2} \ne \delta \Identity$ implies $\myGenVarS_{\underline{i} 1} \ne \myGenVarS_{\underline{i} 2}$ and thus that our problem $S$ was not symmetric with respect to the just eliminated literal and by proposition~\ref{SAT_list_proposition} $S \ne \delta \Identity$ and it is satisfiable.

To prove that $\myGenVarS_{\underline{n} 1} = \myGenVarS_{\underline{n} 2}$ for the last expression of (\ref{formula_implication_list_2}) implies $S = 0$ we start from the last step where we can check easily if $\myGenVarS_{\underline{n} 1} = \myGenVarS_{\underline{n} 2}$ and in both cases (unsatisfiable or tautology) the expression is fully symmetric in all (none actually) the literals and so if $\myGenVarS_{\underline{n} 1} = \myGenVarS_{\underline{n} 2}$ we also have $\myGenVarS_{\underline{n} 1} = \myGenVarS_{\underline{n} 2} = \delta \Identity$ with $\delta \in \{0, 1\}$.

We show that the property of being fully symmetric of the expression at hand propagates ``upstream'' along the implication chain (\ref{formula_implication_list_2}) and when reaches the first element of the chain proofs that $S = \delta \Identity$.

Applying to any $\myGenVarS_{\underline{i} 1} - \myGenVarS_{\underline{i} 2}$ with $n'$ literals recursion (\ref{formula_decomposition_signed_sum}) with respect to literal $\literal_j$ we get, with heavy but hopefully clear notation,
\begin{equation}
\label{formula_implication_list_recursion}
\myGenVarS_{\underline{i} 1} - \myGenVarS_{\underline{i} 2} = \myconjugate{\literal}_j \myGenVarS_{\underline{i j} 1 1} + \literal_j \myGenVarS_{\underline{i j} 1 2} - \myconjugate{\literal}_j \myGenVarS_{\underline{i j} 2 1} - \literal_j \myGenVarS_{\underline{i j} 2 2} = \myconjugate{\literal}_j (\myGenVarS_{\underline{i j} 1 1} - \myGenVarS_{\underline{i j} 2 1}) + \literal_j (\myGenVarS_{\underline{i j} 1 2} - \myGenVarS_{\underline{i j} 2 2})
\end{equation}
and we go at the next right level of (\ref{formula_implication_list}) bringing the expression with $n' - 1$ literals
$$
(\myGenVarS_{\underline{i j} 1 1} - \myGenVarS_{\underline{i j} 2 1}) - (\myGenVarS_{\underline{i j} 1 2} - \myGenVarS_{\underline{i j} 2 2}) = \delta \Identity \dotinformula
$$
If at the next level holds not only this relation but also the stronger relation $(\myGenVarS_{\underline{i j} 1 1} - \myGenVarS_{\underline{i j} 2 1}) = (\myGenVarS_{\underline{i j} 1 2} - \myGenVarS_{\underline{i j} 2 2}) = \delta \Identity$ with $\delta \in \{0, 1\}$ this implies that the two expressions are necessarily fully symmetric in their $n' - 1$ literals, and this implies in turn that also for the expression at the previous step (\ref{formula_implication_list_recursion})
$$
\myGenVarS_{\underline{i} 1} - \myGenVarS_{\underline{i} 2} = (\myconjugate{\literal}_j + \literal_j)(\myGenVarS_{\underline{i j} 1 1} - \myGenVarS_{\underline{i j} 2 1}) = \delta \Identity \qquad \delta \in \{0, 1\}
$$
and thus that $\myGenVarS_{\underline{i} 1} - \myGenVarS_{\underline{i} 2}$ is fully symmetric in all its $n'$ literals. We conclude that this property propagates upstream the implication chain (\ref{formula_implication_list_2}) until it reaches the first level, just after $S$ and, with proposition~\ref{SAT_recursion_algorithm}, $S = 0$.
\end{proof}

This implication chain defines recursive algorithm~\ref{SAT_list_algorithm} that, starting from $S$ eliminates one literal at the time producing a list of sum of signed \SAT{} problems (\ref{formula_signed_sum}) that, by proposition~\ref{SAT_list_proposition}, at the last step can have $r \le 2^n$ \SAT{} problems. At each step we easily see that each $S_j = 0$ can be eliminated from $\myGenVarS$ and that each satisfiable $S_j$, being not fully symmetric, proves that the initial problem $S$ is satisfiable.

We remark that if in the running of the recursive algorithm $\myGenVarS$ contains terms $S_j$ and $- S_j$ they can not be ``simplified'' since if $S_j$ is satisfiable, namely $S_j \ne 0$, replacing $S_j - S_j$ with $0$ would replace a non fully symmetric term with a fully symmetric one, thus hiding a source of asymmetry (while the simplification can be done in a non recursive algorithm). A very preliminary version of the algorithm written in Mathematica solves correctly random \SAT{} problems with tens of literals. A crude analysis of the running time, assuming recursion relation $A(n) = 2 A(n-1)$, gives obviously $A(n) = \bigO{2^{n-1}}$.
\opt{margin_notes}{\mynote{mbh.note: commented a part with other possibilities}}%
%
%but we remark that assuming $A(n) = 2 A((1-\epsilon)(n-1))$, actually found in practice, would give a completely different result.

\newpage %incredible: this newpage goes here to put bibliography in a new page

\begin{algorithm}
\caption{\SAT-list symmetry algorithm}
\label{SAT_list_algorithm}
\begin{algorithmic}[5]
\Require A signed sum of \SAT{} problems $\myGenVarS$ in $n$ literals (\ref{formula_signed_sum})
\Ensure $\mathrm{T}$ if $\myGenVarS \ne \delta \Identity$ ($\delta \in \R$), $\mathrm{F}$ otherwise
\State {}

\State Start: possibly simplify incoming $\myGenVarS$ eliminating all unsatisfiable $S_j$

\If{$\myGenVarS$ is empty}
\State return $\mathrm{F}$ \Comment{$0$ is fully symmetric}
\EndIf

\If{$\myGenVarS$ contains any satisfiable problem}
\State return $\mathrm{T}$ \Comment{a satisfiable problem is not symmetric}
\EndIf

\State choose next literal $\literal_i$ \Comment{this is a delicate point}

\For{any \SAT{} problem $s_j S_{j}$ of $\myGenVarS$}
\State generate \SAT{} problems $S_{j \underline{i} 1}$ and $S_{j \underline{i} 2}$
%\If{$S_{j \underline{i} 0} S_{j \underline{i} 1} - S_{j \underline{i} 0} S_{j \underline{i} 2}$ can be calculated} %\Comment{typically the easily computable result is either $0$ or a tautology}
%\State replace $s_j S_{j}$ with the result of $s_j (S_{j \underline{i} 0} S_{j \underline{i} 1}- S_{j \underline{i} 0} S_{j \underline{i} 2})$
%\Else
\State replace each $s_j S_{j}$ with $s_j (S_{j \underline{i} 1}- S_{j \underline{i} 2})$
%\EndIf
\EndFor
\State {\bf recur} with the updated sum of problems $\myGenVarS$ in $n-1$ literals
\State \Return result of recursion
\end{algorithmic}
\end{algorithm}
\newpage

%\bigskip
%\bigskip
%\bigskip
%\bigskip
%\bigskip
%\bigskip
%\bigskip

\opt{x,std,AACA}{% in all cases - but arXiv & JMP - standard BibTeX bibliography

\bibliographystyle{plain} % or plain or.... see e.g.\ http://amath.colorado.edu/documentation/LaTeX/reference/faq/bibstyles.html#styles
\bibliography{mbh}
}

\opt{arXiv,JMP}{% only for arXiv & JMP we need to include here the L-A-S-T version of file .bbl
%
%
%\begin{thebibliography}{1}

%\end{thebibliography}
%
%
}

\opt{final_notes}{
\newpage

\section*{Things to do, notes, etc.......}

\subsection*{Other propositions, other material, etc.......}
\subsubsection*{Things to do}
\begin{itemize}
\item put the part on NOT before the part on OR and senplify everything;
\item other ideas:
\begin{itemize}
\item as in Onsager solutions find the transformation in $\R^{n,n}$ corresponding to a given $\SAT$ problem in $\myClg{}{}{\R^{n,n}}$
\item exploit better the solution with $\Delta = \Identity$ writing $\Delta = (\literal_i + \myconjugate{\literal}_i) \Delta'$ with $\Delta'$ in $\myClg{}{}{\R^{n-1,n-1}}$; something is done in the alternative proof of proposition~\ref{e_i_invariance} at the end;
\item exploit idea of secular equation;
\item ...
\end{itemize}

\end{itemize}

%\newpage

\section*{Tentative New Appendix}
In section~\ref{unSAT_algorithm} we outlined a \SAT{} algorithm that exploits the relaxation of (\ref{formula_sat_recursion}) to (\ref{formula_e_i_invariance}): we describe it here together with a proof of its correctness.

We start iterating once the projection formula (\ref{formula_SAT_recursive_expansion})
$$
\begin{array}{l l l}
S & = & \myconjugate{\literal}_i S_{\underline{i} 1} + \literal_i S_{\underline{i} 2} = \myconjugate{\literal}_i (\myconjugate{\literal}_j S_{\underline{i j} 1 1} + \literal_j S_{\underline{i j} 1 2}) + \literal_i (\myconjugate{\literal}_j S_{\underline{i j} 2 1} + \literal_j S_{\underline{i j} 2 2}) = \\
& = & \myconjugate{\literal}_i \myconjugate{\literal}_j S_{\underline{i j} 1 1} + \myconjugate{\literal}_i \literal_j S_{\underline{i j} 1 2} + \literal_i \myconjugate{\literal}_j S_{\underline{i j} 2 1} + \literal_i \literal_j S_{\underline{i j} 2 2} = \cdots
\end{array}
$$
where \eg $S_{\underline{i j} 1 1}$ stands for the first of the four \SAT{} problems without literals $\literal_i$ and $\literal_j$. After $k$ iterations on literals $\literal_{i_1}, \literal_{i_2}, \ldots, \literal_{i_k}$ we write
$$
S = \sum_{\alpha_1, \alpha_2, \ldots, \alpha_k} R_{\alpha_1 \alpha_2 \cdots \alpha_k} S_{\underline{i_1 i_2 \cdots i_k} \alpha_1 \alpha_2 \cdots \alpha_k}
$$
where the sum is over $k$ indices $1 \le \alpha_i \le 2$, has $2^k$ terms and $R_{\alpha_1 \alpha_2 \cdots \alpha_k}$ stands for the product of $k$ literals and the indices $\alpha_1 \alpha_2 \cdots \alpha_k$ specify the unique combination of complemented and plain literals, $\alpha_j = 1,2$ corresponding respectively to $\myconjugate{\literal}_{i_j}, \literal_{i_j}$, for example $R_{1 2 1} S_{\underline{i_1 i_2 i_3} 1 2 1} = \myconjugate{\literal}_{i_1} \literal_{i_2} \myconjugate{\literal}_{i_3} S_{\underline{i_1 i_2 i_3} 1 2 1}$. We can extend proposition~\ref{e_i_invariance} to this case:
\begin{MS_Proposition}
\label{boh}
A non empty $\SAT$ problem $S$ is invariant with respect to $\literal_{i_k}$, namely $\mygen_{2 {i_k} - 1} \; S \; \mygen_{2 {i_k} - 1} = S$, if and only if are satisfied the $2^{k-1}$ relations
\begin{equation}
\label{formula_SAT_recursive_expansion_k_condition}
S_{\underline{i_1 i_2 \cdots i_{k - 1}} \alpha_1 \alpha_2 \cdots \alpha_{k - 1} 1} = S_{\underline{i_1 i_2 \cdots i_{k - 1}} \alpha_1 \alpha_2 \cdots \alpha_{k - 1} 2}
\end{equation}
for all $2^{k-1}$ values taken by indices $\alpha_1, \alpha_2, \ldots, \alpha_{k - 1}$.
\end{MS_Proposition}
\begin{proof}
For $k = 1$ this is proposition~\ref{e_i_invariance}; to verify its validity in the generic $k$ case we write $S$ summing only over $k-1$ indices and making explicit the $k$-th sum
$$
S = \sum_{\alpha_1, \alpha_2, \ldots, \alpha_{k - 1}} R_{\alpha_1 \alpha_2 \cdots \alpha_{k - 1}} (\myconjugate{\literal}_{i_k} S_{\underline{i_1 i_2 \cdots i_{k - 1}} \alpha_1 \alpha_2 \cdots \alpha_{k - 1} 1} + \literal_{i_k} S_{\underline{i_1 i_2 \cdots i_{k - 1}} \alpha_1 \alpha_2 \cdots \alpha_{k - 1} 2})
$$
and supposing that (\ref{formula_SAT_recursive_expansion_k_condition}) holds we obtain
$$
S = \sum_{\alpha_1, \alpha_2, \ldots, \alpha_{k - 1}} R_{\alpha_1 \alpha_2 \cdots \alpha_{k - 1}} (\myconjugate{\literal}_{k} + \literal_{k}) S_{\underline{i_1 i_2 \cdots i_{k - 1}} \alpha_1 \alpha_2 \cdots \alpha_{k - 1} 1}
$$
that is manifestly invariant for the exchange of $\literal_{k}$ with $\myconjugate{\literal}_{k}$.
\end{proof}

\subsection*{Old parts, removed, unnecessary, etc.......}

\subsubsection*{Old first part of section: An algorithm to test for unsatisfiability}
\label{old_unSAT_algorithm}
With the result of previous paragraph we outline a \SAT{} algorithm. The basic idea is that if a given \SAT{} problem is either unsatisfiable or a tautology it has to have the maximally symmetric form (\ref{formula_identity_def}) of a scalar element of $\myClg{}{}{\R^{n,n}}$, namely $\delta \Identity$ with $\delta \in \R$. If any of the symmetry relations of theorem~\ref{SAT_unsat_thm} fails the problem is necessarily satisfiable.

To prove some simple results we briefly reproduce the well known Jordan Wigner construction for the matrix form of Clifford algebra: the purpose is twofold on one side make as simple as possible some properties of Clifford algebra and on the other show convincingly that even if we deal with matrices of dimension $2^n$, and in $\SAT$ problems one is interested in the limit $n \to \infty$, we can manipulate them manipulating just $n$ matrices $2 \times 2$.

As already stated $\myClg{}{}{\R^{n,n}} \myisom \R(2^n)$ and we can build explicitly this isomorphism (technically we fix the injection of the vector space in the isomorphic matrix algebra $\beta: V \to \R(2^n)$). Starting from $n = 1$, when $\myClg{}{}{\R^{1,1}} \myisom \R(2)$, we choose (it's not the only possible choice)
$$
\mygen_1 = \left(\begin{array}{r r} 0 & 1 \\ 1 & 0 \end{array}\right) := \sigma \qquad \mygen_2 = \left(\begin{array}{r r} 0 & -1 \\ 1 & 0 \end{array}\right) := \epsilon \qquad \mygen_1 \mygen_2 = \left(\begin{array}{r r} 1 & 0 \\ 0 & -1 \end{array}\right) := \tau
$$
and from this choice we can build, exploiting $\myClg{}{}{\R^{n,n}}$ properties, the recursion relation that, from the $2 n$ generators of $\myClg{}{}{\R^{n,n}}$, gives the $2 n + 2$ of $\myClg{}{}{\R^{n+1,n+1}}$, namely, indicating tensor product with $\otimes$,
\begin{equation*}
\label{old_formula_space_signature}
\left\{ \begin{array}{l}
\mygen_i \otimes \tau \qquad 1 \le i \le 2 n \\
\Identity_{2^n} \otimes \sigma \\
\Identity_{2^n} \otimes \epsilon
\end{array} \right.
\end{equation*}
where $\Identity_{2^n}$ is the identity matrix of $\R(2^n)$. For our purposes it is instructive to write the generators explicitly in the generic $n$ case
\begin{equation}
\label{old_formula_generators_matrices}
\begin{array}{r@{} l l}
\mygen_1 & {}= & \boldsymbol{\sigma} \otimes \tau \otimes \tau \otimes \tau \otimes \tau \otimes \tau \otimes \cdots \otimes \tau \\
\mygen_2 & {}= & \boldsymbol{\epsilon} \otimes \tau \otimes \tau \otimes \tau \otimes \tau \otimes \tau \otimes \cdots \otimes \tau \\
\mygen_3 & {}= & \Identity \otimes \boldsymbol{\sigma} \otimes \tau \otimes \tau \otimes \tau \otimes \tau \otimes \cdots \otimes \tau \\
\mygen_4 & {}= & \Identity \otimes \boldsymbol{\epsilon} \otimes \tau \otimes \tau \otimes \tau \otimes \tau \otimes \cdots \otimes \tau \\
& & \qquad \cdots \\
\mygen_{2 i - 1} & {}= & \Identity \otimes \Identity \otimes \cdots \otimes \Identity \otimes \boldsymbol{\sigma} \otimes \tau \otimes \cdots \otimes \tau \\
\mygen_{2 i} & {}= & \Identity \otimes \Identity \otimes \cdots \otimes \Identity \otimes \boldsymbol{\epsilon} \otimes \tau \otimes \cdots \otimes \tau\\
& & \qquad \cdots \\
\mygen_{2 n - 1} & {}= & \Identity \otimes \Identity \otimes \Identity \otimes \Identity \otimes \Identity \otimes \cdots \otimes \Identity \otimes \boldsymbol{\sigma} \\
\mygen_{2 n} & {}= & \Identity \otimes \Identity \otimes \Identity \otimes \Identity \otimes \Identity \otimes \cdots \otimes \Identity \otimes \boldsymbol{\epsilon}\\
\end{array}
\end{equation}
where there are $n-1$ tensor products in each row and we marked in bold matrices $\sigma$ and $\epsilon$ for visibility, moreover here $\Identity$ stands for the identity matrix in $\R(2)$. It is immediate to verify that these matrices satisfy the commutation relations (\ref{formula_generators}) of $\myClg{}{}{\R^{n,n}}$ and that $\mygen_{i}^2 = (-1)^{i + 1} \Identity$.

Let us now see the form of a generic literal, $\literal_i$, that we may write as $\literal_i \prod_{\stackrel{j = 1}{j \ne i}}^n \anticomm{q_j}{p_j}$. For $n = 1$ it is easy to see that $qp = \frac{1}{2} (\Identity + \tau)$, $pq = \frac{1}{2} (\Identity - \tau)$ and that $qp + pq = \Identity$. It is then easy to verify that for generic $n$ a literal takes the form
\begin{equation}
\label{old_formula_generic_literal}
\literal_i = \frac{1}{2} \Identity \otimes \Identity \otimes \cdots \otimes \Identity \otimes (\Identity \pm \tau) \otimes \Identity \otimes \cdots \otimes \Identity
\end{equation}
where $(\Identity \pm \tau)$ appears at $i$-th position and it is easy to see that, as anticipated, that literals can be represented by diagonal matrices. The generic term of the product (\ref{formula_SAT_EFB_2}) is
$$
\Identity - z_j = \Identity - \myconjugate{\literal}_r \myconjugate{\literal}_s \cdots \myconjugate{\literal}_t = \Identity \otimes \Identity \otimes \cdots \otimes \Identity \; - \; \frac{1}{2^k} \Identity \otimes \cdots \otimes (\Identity \pm \tau) \otimes \cdots \otimes (\Identity \pm \tau) \otimes \cdots \otimes (\Identity \pm \tau) \otimes \cdots
$$
where the $k$ literals $\myconjugate{\literal}_r, \myconjugate{\literal}_s, \ldots, \myconjugate{\literal}_t$ appear in the direct products as $(\Identity \pm \tau)$ at positions $r, s, \ldots, t$ and it is also manifest that also this element of $\mysetS$ is represented by a diagonal matrix with $0, 1$ on the diagonal.

In this matrix algebra the primitive idempotents forming the orthonormal basis of $\myAbelsa$ (\ref{formula_identity_def}) are the $2^n$ diagonal matrices
$$
(\Identity \pm \tau) \otimes (\Identity \pm \tau) \otimes \cdots \otimes (\Identity \pm \tau) \dotinformula
$$

To apply the invariance test $\mygen_i z_j \mygen_i^{-1}$ and looking at the matrix formulation (\ref{old_formula_generators_matrices}) we easily see that we do not need to manipulate matrices of $\R(2^n)$ but that the test reduces to testing $n$ copies of $\R(2)$ matrices. Let us examine first the case $n = 1$, when $\myClg{}{}{\R^{1,1}} \myisom \R(2)$ where
$$
\sigma \tau \sigma^{-1} = \epsilon \tau \epsilon^{-1} = - \tau \qquad \qquad \tau \tau \tau^{-1} = \tau
$$
and thus, for the generic literal of this case $\literal = \frac{1}{2} (\Identity \pm \tau)$,
$$
\sigma \literal \sigma^{-1} = \epsilon \literal \epsilon^{-1} = \myconjugate{\literal} \qquad \qquad \tau \literal \tau^{-1} = \literal
$$
and we notice that to test invariance it is sufficient to test invariance for $\sigma$ since it is equivalent to invariance for $\epsilon$. In this simple case it is easy to verify also theorem~\ref{SAT_unsat_thm} since the only unsatisfiable \SAT{} for $n = 1$, $\literal \myconjugate{\literal}$, is actually invariant since $\sigma \literal \myconjugate{\literal} \sigma^{-1} = \literal \myconjugate{\literal}$.

Going to the generic $n$ case also here it will be sufficient to test invariance just for the odd generators $\mygen_{2 i - 1}$ for which $\mygen_{2 i - 1}^{-1} = \mygen_{2 i - 1}$ and with (\ref{old_formula_generators_matrices}) and (\ref{old_formula_generic_literal}) it is simple to see that
\begin{equation}
\label{old_formula_literal_invariances}
\mygen_{2 i - 1} \literal_i \mygen_{2 i - 1} = \myconjugate{\literal}_i \;\; \mbox{for} \;\; 1 \le i \le n \qquad \qquad \mygen_{2 i - 1} \literal_j \mygen_{2 i - 1} = \literal_j \;\; \mbox{for} \;\; i \ne j
\end{equation}
so the literal $\literal_i$ is complemented by ``its'' generator $\mygen_{2 i - 1}$ and left invariant by all other generators, a typical case of reflections in Clifford algebras.

This sounds legitimate also in the purely logical formulation of \SAT{} problems: a problem is unsatisfiable if and only if exchanging any literal for its complement the result is not altered.

\bigskip

To derive an actual algorithm from these relations we need a technical
\begin{MS_lemma}
\label{old_SAT_recursive_expansion}
Given a non empty $k\SAT$ problem $S$ with $n$ literals, for any literal $\literal_i$ we may write $S$ as
\begin{equation}
\label{old_formula_SAT_recursive_expansion}
S = \myconjugate{\literal}_i S_0 S_1 + \literal_i S_0 S_2
\end{equation}
\opt{margin_notes}{\mynote{mbh.note: here $+$ coincides with the logical OR}}%
where $S_0$ is a $k\SAT$ problem in $n - 1$ literals (all but the $i$-th) and $S_1$ and $S_2$ are $(k-1)\SAT$ problems in $n - 1$ literals, all defined in the proof; moreover all elements of this relation are in $\mysetS$.
\end{MS_lemma}

\begin{proof}
We separate the clauses of $S$ into three classes: those in which $\literal_i$ appears, those in which $\myconjugate{\literal}_i$ appears and the others in which neither appear. Obviously these three classes form three ``reduced'' $k\SAT{}$ problems that we call respectively $S_1'$, $S_2'$ and $S_0$, clearly $S = S_0 S_1' S_2'$.

Let us now consider the clauses of the first set $S_1'$, those with $\literal_i$, that we put in form (\ref{formula_SAT_EFB_3}). More precisely, since all these clauses contain $\literal_i$, when written in the form $\Identity - z_j$ certainly $\myconjugate{\literal}_i$ appears in all $z_j$ and so for these clauses the form (\ref{formula_SAT_EFB_3}) is
$$
S_1' = \Identity - \myconjugate{\literal}_i \Delta_1
$$
with $\Delta_1$ ``free'' of literals $\myconjugate{\literal}_i$. Similarly $S_2'$ takes the form $S_2' = \Identity - \literal_i \Delta_2$ and we may write our original \SAT{} problem $S$
$$
%\label{formula_sat_subproblems}
S = S_0 (\Identity - \myconjugate{\literal}_i \Delta_1) (\Identity - \literal_i \Delta_2) = S_0 \left(\Identity - \myconjugate{\literal}_i \Delta_1 - \literal_i \Delta_2\right)
$$
and, defining and substituting $S_l := \Identity - \Delta_l, l = 1,2$, we get with (\ref{old_formula_complemented_assignement_one_literal})
$$
S = S_0 \left(\Identity - \literal_i - \myconjugate{\literal}_i + \myconjugate{\literal}_i S_1 + \literal_i S_2 \right) = \myconjugate{\literal}_i S_0 S_1 + \literal_i S_0 S_2 \dotinformula
$$
It is manifest that $S_0$ is a $k\SAT$ problem in $n - 1$ literals (all but the $i$-th); for the other two problems this can be seen remembering the definition of \eg $S_1' = \Identity - \myconjugate{\literal}_i \Delta_1$; it is a simple exercise to show that taking the very same clauses that form $S_1'$ and \emph{removing} from them $\literal_i$ we obtain a $(k-1)\SAT$ problem in $n - 1$ literals whose form (\ref{formula_SAT_EFB_3}) is $S_1 = \Identity - \Delta_1$. Since all problems in this relation are $\SAT$ problems they are in $\mysetS$; in particular $\literal_i S_0 S_2$ and $\myconjugate{\literal}_i S_0 S_1$ belong respectively to the two orthogonal subspaces $\literal_i \myAbelsa$ and $\myconjugate{\literal}_i \myAbelsa$ of $\myAbelsa$ (\ref{formula_P_subspaces}) being precisely the two projections of $S$ in these subspaces.
\end{proof}

\subsubsection*{Old section: Satisfiability in Clifford algebra}
\label{old_SAT_reformulation}
Given a SAT problem with $m > 0$ clauses and $n$ logical variables $x_i$ we consider the Clifford algebra $\myClg{}{}{\R^{n,n}}$ that is isomorphic to the algebra of real matrices $\R(2^n)$. This algebra is more easily manipulated exploiting the properties of its Extended Fock Basis (EFB, see \cite{Budinich_2016} and references therein) with which any algebra element is a linear superposition of simple spinors.

We remind that the $2 n$ generators of the algebra $\mygen_{i}$ form an orthonormal basis of the neutral vector space $V = \R^{n,n}$ with \eg
\begin{equation}
\label{old_formula_generators}
\mygen_i \mygen_j + \mygen_j \mygen_i := \anticomm{\mygen_i}{\mygen_j} = 2 \delta_{i j} (-1)^{i+1} \qquad i,j = 1,2, \ldots, 2 n
\end{equation}
while the Witt, or null, basis of $V$ is:
\begin{equation}
\label{old_formula_Witt_basis}
\left\{ \begin{array}{l l l}
p_{i} & = & \frac{1}{2} \left( \mygen_{2i-1} + \mygen_{2i} \right) \\
q_{i} & = & \frac{1}{2} \left( \mygen_{2i-1} - \mygen_{2i} \right)
\end{array} \right.
%\Rightarrow
%\left\{\begin{array}{l l l}
%\mygen_{2i-1} & = & p_{i} + q_{i} \\
%\mygen_{2i} & = & p_{i} - q_{i}
%\end{array} \right.
\quad i = 1,2, \ldots, n
\end{equation}
that, with $\mygen_{i} \mygen_{j} = - \mygen_{j} \mygen_{i}$, gives
\begin{equation}
\label{old_formula_Witt_basis_properties}
\anticomm{p_{i}}{p_{j}} = \anticomm{q_{i}}{q_{j}} = 0
\qquad
\anticomm{p_{i}}{q_{j}} = \delta_{i j}
\end{equation}
showing that all $p_i, q_i$ are mutually orthogonal, also to themselves, that implies $p_i^2 = q_i^2 = 0$, at the origin of the name ``null'' given to these vectors. Simple spinors forming EFB are products of $n$ or more null vectors (\ref{old_formula_Witt_basis}).

To represent \SAT{} problems we will assume that $\mathrm{F}$ is 0 and $\mathrm{T}$ any non zero element. Moreover we will not need the full algebra $\myClg{}{}{\R^{n,n}}$ but we will restrict to the even, Abelian, subalgebra $\myAbelsa$ given by the vector space spanned by the $2^n$ primitive idempotents $\myprimidemp_i$ of the algebra. We remind the standard properties of the primitive idempotents
\begin{equation}
\label{old_formula_primitive_idempotents}
\myprimidemp_i^2 = \myprimidemp_i \quad (\Identity - \myprimidemp_i)^2 = \Identity - \myprimidemp_i \quad \myprimidemp_i (\Identity - \myprimidemp_i) = 0 \quad \myprimidemp_i \myprimidemp_j = \delta_{i j} \myprimidemp_i
\end{equation}
and, since $\myClg{}{}{\R^{n,n}}$ is a simple algebra, the unit element of the algebra $\Identity$ can be expressed as the sum of its primitive idempotents
\begin{equation}
\label{old_formula_identity_def}
\Identity = \sum_{i = 1}^{2^n} \myprimidemp_i = \anticomm{q_1}{p_1} \anticomm{q_2}{p_2} \cdots \anticomm{q_n}{p_n} = \prod_{j = 1}^{n} \anticomm{q_j}{p_j}
\end{equation}
where the second form, a product of $n$ anticommutators, is the expression of identity in EFB, and the full expansion of these anticommutators generate $2^n$ terms each term being one of the primitive idempotents (and a simple spinor). In the isomorphic matrix algebra $\R(2^n)$ $\myAbelsa$ is usually (but not necessarily) the subalgebra of diagonal matrices and the primitive idempotents are the $2^n$ diagonal matrices with a single $1$ on the diagonal.

The subalgebra $\myAbelsa$ contains a subset
\begin{equation}
\label{old_formula_S_def}
\mysetS = \left\{ \sum_{i = 1}^{2^n} \delta_i \myprimidemp_i : \delta_i \in \left\{0, 1 \right\} \right\} \subset \myAbelsa
\end{equation}
that is closed under multiplication but not under addition and is thus not even a subspace. With primitive idempotent properties (\ref{old_formula_primitive_idempotents}) it is simple to prove
\opt{margin_notes}{\mynote{mbh.note: the proof is commented. Is in general any idempotent a sum of primitive idempotents ?}}%
\begin{MS_Proposition}
\label{old_S_properties}
$s \in \mysetS$ if and only if $s$ is an idempotent, $s^2 = s$
\end{MS_Proposition}
%
%\begin{proof}
%Let $s \in \mysetS$, by primitive idempotent properties (\ref{old_formula_primitive_idempotents}) follows immediately that $s^2 = s$. Conversely let $s^2 = s$; since $s$ is an idempotent it must be in $\myAbelsa$ (this is to be verified); let $s = \sum_{i = 1}^{2^n} \gamma_i \myprimidemp_i$, by primitive idempotent properties (\ref{old_formula_primitive_idempotents}) it follows that $s^2 = \sum_{i = 1}^{2^n} \gamma_i^2 \myprimidemp_i$ thus the hypothesis implies that $\gamma_i^2 = \gamma_i$ that in $\R$ (?) holds only if $\gamma_i \in \{0, 1\}$.
%\end{proof}
%
\noindent and thus $\mysetS$ is the set of the idempotents, in general not primitive; a simple consequence is that for any $s \in \mysetS$ also $(\Identity - s) \in \mysetS$. We remark also that the $2^n$ primitive idempotents $\myprimidemp_i$ form an orthonormal basis of $\myAbelsa$ as vector space and provide also a unique identification of all elements of $\mysetS$.

To represent Boolean variables in the subalgebra of $\myClg{}{}{\R^{n,n}}$ we map literal $\literal_i \to q_i p_i$ and $\myconjugate{\literal}_i \to p_i q_i$ and with Witt basis properties (\ref{old_formula_Witt_basis_properties}) we have
$$
q_i p_i \, q_i p_i = q_i p_i \quad q_i p_i \, p_i q_i = 0 \quad q_i p_i \, q_j p_j = q_j p_j \, q_i p_i \quad q_i p_i + p_i q_i = 1
$$
that, mapping the logical AND and OR to Clifford product and sum, can be read as the logical relations for literals
$$
\literal_i \land \literal_i \equiv \literal_i \quad \literal_i \land \myconjugate{\literal}_i \equiv \mathrm{F} \quad \literal_i \land \literal_j \equiv \literal_j \land \literal_i \quad \literal_i \lor \myconjugate{\literal}_i \equiv \mathrm{T}
$$
where $\equiv$ represents the logical equivalence of the expressions, namely that for all possible values taken by the literals the two expressions are equal.

By proposition~\ref{old_S_properties} and $q_i p_i \, q_i p_i = q_i p_i$ descends that all the literals mapped in Clifford algebra are in $\mysetS$; it is nevertheless instructive to derive this property directly from EFB formalism applied to \eg $q_1 p_1$, with (\ref{old_formula_identity_def})
$$
q_1 p_1 = q_1 p_1 \Identity = q_1 p_1 \anticomm{q_2}{p_2} \cdots \anticomm{q_n}{p_n}
$$
since $q_1 p_1 \anticomm{q_1}{p_1} = q_1 p_1$ (and more in general $\literal_i \anticomm{q_i}{p_i} = \literal_i (\literal_i + \myconjugate{\literal}_i) = \literal_i$); the full expansion is thus a sum of $2^{n - 1}$ EFB terms that are primitive idempotents and thus $q_1 p_1 \in \mysetS$. From the logical viewpoint this can be interpreted as the property that when $\literal_1$ is fixed the other, unspecified, $n-1$ literals $\literal_2, \ldots, \literal_n$ are free to take all possible $2^{n - 1}$ values.

Since the set $\mysetS$ is closed under multiplication all product of literals, corresponding to logical AND, are $1\SAT$ formulas that are in $\mysetS$ and thus are idempotents. The generalization of previous formula to the case of a product of an arbitrary number of literals is
$$
\literal_r \literal_s \cdots \literal_t = \literal_r \literal_s \cdots \literal_t \Identity = \literal_r \literal_s \cdots \literal_t \prod_{\stackrel{i = 1}{i \ne r, s, \ldots, t}}^n \anticomm{q_i}{p_i} \dotinformula
$$

At this point it should be manifest that the $2^n$ primitive idempotents $\myprimidemp_i$ are in one to one correspondence with the possible $2^n$ assignments to the $n$ literals $\literal_i$, for example given an assignment of the $n$ literals one finds
$$
\literal_1 \myconjugate{\literal}_2 \cdots \literal_n \to q_1 p_1 p_2 q_2 \cdots q_n p_n
$$
that is one of the primitive idempotents of (\ref{old_formula_identity_def}). A curiosity is that, by EFB properties \cite{Budinich_2016}, converting the literals to a binary string with substitutions $\literal \to 0, \myconjugate{\literal} \to 1$ and reading this string as an integer in base 2 we obtain the (equal) row and column indices giving the position of the corresponding idempotent on the diagonal of the matrix of $\myClg{}{}{\R^{n,n}}$.

A property that we will use in the sequel is that, for every $1 \le i \le n$, the vector space $\myAbelsa$ may be written as a direct sum of orthogonal subspaces (and subalgebras)
\begin{equation}
\label{old_formula_P_subspaces}
\myAbelsa = \myAbelsa_i \oplus \myconjugate{\myAbelsa}_i = q_i p_i \myAbelsa \oplus p_i q_i \myAbelsa
\end{equation}
\opt{margin_notes}{\mynote{mbh.ref: for a proof see log p. 695}}%
where $\myAbelsa_i$ and $\myconjugate{\myAbelsa}_i$ are subspaces of dimension $2^{n - 1}$ that can be obtained by projections $\myAbelsa_i = q_i p_i \myAbelsa$ as is simple to see writing the vector expansion of any vector of $\myAbelsa$ in the idempotent basis and remembering that $q_i p_i \anticomm{q_i}{p_i} = q_i p_i$. Clearly both subspaces contain corresponding orthogonal subsets $\mysetS_i$ and $\myconjugate{\mysetS}_i$ and, by (\ref{old_formula_S_def}), for any $s \in \mysetS_i$ then $\Identity - s \in \myconjugate{\mysetS}_i$.

When dealing with logical expressions we must distinguish two cases in which literals may appear: a product of literals may be given the meaning of an assignment of values to logical variables, for example we may interpret $\literal_i \literal_j \myconjugate{\literal}_k$ as $\literal_i \equiv \literal_j = \mathrm{T}, \literal_k = \mathrm{F}$, or as a $1\SAT$ formula with unassigned literals. With our mapping in Clifford algebra this duality is respected and we can test if an assignment, \eg $\literal_i \literal_j \myconjugate{\literal}_k$, satisfies a given formula, \eg $\literal_i \myconjugate{\literal}_k$ ``substituting'' the assignment in the formula, that in Clifford algebra is performed simply calculating their product: if the result is not zero, $\mathrm{T}$, this means that the given assignment satisfies the formula, in our example
$$
\literal_i \myconjugate{\literal}_k \; \; \literal_i \literal_j \myconjugate{\literal}_k = \literal_i \literal_j \myconjugate{\literal}_k \ne 0 \to \mathrm{T}
$$
in this sense we can read $\literal_i \myconjugate{\literal}_i = 0$ as the fact that $\literal_i = \mathrm{F}$ does not satisfy the formula $\literal_i$ and $\literal_i \literal_i = \literal_i$ as the fact that $\literal_i = \mathrm{T}$ satisfies formula $\literal_i$.

We come now to the representation in $\myClg{}{}{\R^{n,n}}$ of the logical OR: unfortunately $\mysetS$ is not closed under addition and so in general the sum of literals belong to the subalgebra $\myAbelsa$ but not to $\mysetS$, for example $\literal_i + \literal_i = 2 \literal_i$ or $(\literal_i + \literal_j)^2 = \literal_i + \literal_j + 2 \literal_i \literal_j$ and we realize that the sum in Clifford algebra does not reproduce faithfully the properties of the logical OR. Still the sum will turn out to be useful to map the logical OR of literals. For example we know that logical clause $\literal_i + \literal_j$ is satisfied by all assignment but $\literal_i \equiv \literal_j = \mathrm{F}$ and this result is reproduced in Clifford algebra where
$$
(q_i p_i + q_j p_j) p_i q_i p_j q_j = 0
$$
whereas all other possible assignments ($q_i p_i p_j q_j, p_i q_i q_j p_j, q_i p_i q_j p_j$) give non zero results and satisfy the formula. It is simple to verify that also
\begin{equation}
\label{old_formula_SAT_in_P_ambiguity}
(\delta_i q_i p_i + \delta_j q_j p_j) p_i q_i p_j q_j = 0 \qquad \forall \delta_i, \delta_j \in \R
\end{equation}
whatever the values taken by the real coefficients. This example shows constructively that the field coefficient that may appear in the sums are irrelevant as far as the mapping of a Boolean expression is concerned. So we can guess that even if $\mysetS$ is not closed under addition we are allowed to use the sum to represent the logical OR of literals.
\opt{margin_notes}{\mynote{mbh.note: old footnote: one could be tempted to use a ``projection'' in $\mysetS$ of the sum assuming that all non zero coefficients are brought to $1$ but this would still not be satisfactory since while from the logical viewpoint $(\literal_i + \literal_j) (\literal_i + \literal_j) = \literal_i + \literal_j$ even with this ``new sum'' we would get $(\literal_i + \literal_j) (\literal_i + \literal_j) = \literal_i + \literal_j + \literal_i \literal_j$.}}%
That this is actually so, at least for the case of \SAT{} problems, is proved by
\begin{MS_Proposition}
\label{old_SAT_in_Cl}
A given SAT problem (\ref{formula_SAT_std}) admits solution if and only if, the corresponding algebraic expression of $\myClg{}{}{\R^{n,n}}$
\begin{equation}
\label{old_formula_SAT_EFB}
S = (\literal_h + \literal_j + \cdots \literal_k) (\literal_l + \literal_o + \cdots + \literal_p) \cdots (\literal_t + \literal_u + \cdots \literal_z) \qquad \literal_i \in \{q_i p_i, p_i q_i \}
\end{equation}
is different from $0$.
\end{MS_Proposition}

\begin{proof}
the result is immediate observing that by distributivity of $\lor$ and $+$ over multiplication in both (\ref{formula_SAT_std}) and (\ref{old_formula_SAT_EFB}), together with the additional rules $\literal_i \literal_i = \literal_i$ and $\literal_i \myconjugate{\literal}_i = 0$ that again hold in both cases, the expansion of the two formulas are identical and that non zero terms, in both cases, identify all and only the assignments of literals that satisfy the \SAT{} formula (\ref{formula_SAT_std}).
\opt{margin_notes}{\mynote{mbh.note: tricky point, actually in the algebraic expansion some terms make have coefficients $> 1$ for repeated terms whereas in logical expansion all coefficients are $1$}}%
\end{proof}

With this proposition we have transformed a \SAT{} problem in an equivalent calculation in subalgebra $\myAbelsa$ of $\myClg{}{}{\R^{n,n}}$ but to proceed we need a better encoding for (\ref{old_formula_SAT_EFB}). To do this we deal with the case of the logical NOT operator; by Witt basis properties (\ref{old_formula_Witt_basis_properties})
$$
q_i p_i + p_i q_i = 1
$$
that in our setting we can read as $\literal_i + \myconjugate{\literal}_i = 1$. We thus can guess that $\myconjugate{\literal}_i = 1 - \literal_i$ and since in $\myClg{}{}{\R^{n,n}}$ to $1$ we associate $\Identity$
\begin{equation}
\label{old_formula_complemented_assignement_one_literal}
\myconjugate{\literal}_i = \Identity - \literal_i
\end{equation}
that turns out to be a particular case of the more general%
\opt{margin_notes}{\mynote{mbh.note: remark that there exist many idempotents annihilating $\literal_r \literal_s \cdots \literal_t$ but only one is its complement $\Identity - \literal_r \literal_s \cdots \literal_t$}}%
\begin{MS_Proposition}
\label{old_complementary_assignment}
Given any logical assignment, namely $1\SAT$ formula, $\literal_r \literal_s \cdots \literal_t$, its logical complement is
\begin{equation}
\label{old_formula_complemented_assignement}
\myconjugate{\literal_r \literal_s \cdots \literal_t} \equiv \Identity - \literal_r \literal_s \cdots \literal_t
\end{equation}
%$\myconjugate{\literal_r \literal_s \cdots \literal_t} = \Identity - \literal_r \literal_s \cdots \literal_t$.
\end{MS_Proposition}

\begin{proof}
By idempotents properties we obtain immediately $(\Identity - \literal_r \literal_s \cdots \literal_t) \literal_r \literal_s \cdots \literal_t = 0$ and the proof is completed verifying that $\literal_r \literal_s \cdots \literal_t$ is the only assignment of the literals giving this result.
\end{proof}

We are now ready to give a better form to \SAT{} problems in $\myClg{}{}{\R^{n,n}}$

%and it is simple to prove this relation since the formula $\myconjugate{\literal}_i$ is made $\mathrm{F}$ by the only assignment $\literal_i$ since $\myconjugate{\literal}_i \literal_i = (\Identity - \literal_i) \literal_i = 0$. Since $\literal_i \in \mysetS$ we get that also $\myconjugate{\literal}_i \in \mysetS$. This relation generalizes easily to any product of literals to
%\begin{equation}
%\label{old_formula_complemented_assignement}
%\myconjugate{\literal_r \literal_s \cdots \literal_t} = \Identity - \literal_r \literal_s \cdots \literal_t
%\end{equation}
%that can be proved observing that
%$$
%(\Identity - \literal_r \literal_s \cdots \literal_t) \literal_r \literal_s \cdots \literal_t = 0
%$$
%and

\begin{MS_Proposition}
\label{old_SAT_in_Cl_2}
A given SAT problem (\ref{formula_SAT_std}) admits solution if and only if, the corresponding algebraic expression of $\myClg{}{}{\R^{n,n}}$
\begin{equation}
\label{old_formula_SAT_EFB_2}
S = \prod_{j = 1}^m C_j := \prod_{j = 1}^m (\Identity - z_j) \ne 0
\end{equation}
where we define $z_j$ as the product of the complemented literals of clause $C_j$.
\end{MS_Proposition}

\begin{proof}
With proposition~\ref{old_SAT_in_Cl} we just need to prove that from a logical viewpoint each clause $C_j$ is equivalent to $\Identity - z_j$ that, in our setting, is equivalent to show that $C_j = 0$ if and only if $\Identity - z_j = 0$. With De Morgan's relations and (\ref{old_formula_complemented_assignement}) we can write clause $C_j$
$$
C_j = (\literal_r \lor \literal_s \lor \cdots \lor \literal_t) \equiv \myconjugate{\myconjugate{(\literal_r \lor \literal_s \lor \cdots \lor \literal_t)}} \equiv \myconjugate{\myconjugate{\literal}_r \myconjugate{\literal}_s \cdots \myconjugate{\literal}_t} \equiv \Identity - \myconjugate{\literal}_r \myconjugate{\literal}_s \cdots \myconjugate{\literal}_t := \Identity - z_j \dotinformula
$$
\end{proof}

We remark that while with the first representation (\ref{old_formula_SAT_EFB}) $C_j$ in general does not belong to $\mysetS$ in the representation (\ref{old_formula_SAT_EFB_2}) both $z_j$ and $\Identity - z_j$ belong to $\mysetS$ and thus also $S$. Expanding the product of two clauses we get
\begin{equation}
\label{old_formula_z_j_expansion}
(\Identity - z_j) (\Identity - z_k) = \Identity - z_j - z_k + z_j z_k
\end{equation}
and $z_j z_k = 0$ if and only if in $z_j$ and $z_k$ appears the same literal in opposite form ($\literal_i \myconjugate{\literal}_i = 0$). In any way this product is always in $\mysetS$ even if the generic terms of the expansion are not in general in $\mysetS$, \eg $- z_j$. In general the product of $m$ clauses will produce at most $2^m$ terms in the full expansion the first of which is certainly $\Identity$ so that, calling $\Delta$ the other terms we can rewrite (\ref{old_formula_SAT_EFB_2}) as
\begin{equation}
\label{old_formula_SAT_EFB_3}
S = \prod_{j = 1}^m (\Identity - z_j) := \Identity - \Delta
\end{equation}
and in this formulation $S \in \mysetS$ and thus also $\Delta = \Identity - S$ is in $\mysetS$ and all elements of this relation are idempotents.

We remark that proposition~\ref{old_SAT_in_Cl_2} with forms (\ref{old_formula_SAT_EFB_2}) and (\ref{old_formula_SAT_EFB_3}) is a stronger result than that of proposition~\ref{old_SAT_in_Cl} since while the latter represents a \SAT{} problem in $\myAbelsa$ and, as shown in (\ref{old_formula_SAT_in_P_ambiguity}), there are infinite possible representations, the representation in $\mysetS$ is unique by standard basis properties of vector spaces. Moreover only the elements of $\mysetS$ are idempotents.

At this point, since every Boolean expression can be put in the conjunctive normal form of (\ref{formula_SAT_std}), we have the more general result
\begin{MS_Proposition}
\label{old_logical_formulas_in_S}
any Boolean expression $L$ can be represented by $l \in \mysetS$ and $\myconjugate{L}$ by $\Identity - l$ and both are idempotents of $\myClg{}{}{\R^{n,n}}$. Moreover given another Boolean expression $K$ the logical equivalence $L \equiv K$ corresponds to the algebraic equality $l = k$ of their respective representative idempotents in $\mysetS$.
\end{MS_Proposition}
\begin{proof}
To prove the proposition it is sufficient to go to the proof of proposition~\ref{old_SAT_in_Cl} and observe that any Boolean expression in conjunctive normal form is equivalent to a \SAT{} problem and that the expansion of its logical form (\ref{formula_SAT_std}) gives all assignments of the logical variables that satisfy the formula and that these assignments are in one to one correspondence with the expansion of \SAT{} (\ref{old_formula_SAT_EFB_2}) namely $l \in \mysetS$. It is then simple to extend the proof of proposition~\ref{old_complementary_assignment} to the general case of $\myconjugate{L}$ again identifying $L$ with $l \in \mysetS$ written as in (\ref{old_formula_S_def}).
\end{proof}

We conclude this part observing that a logical system is fully defined when are defined, as in our case, the logical AND and the logical NOT. We can exploit this property to get the correct expression of the logical OR in $\myClg{}{}{\R^{n,n}}$. From De Morgan's relation between logical variables we get the corresponding relation in $\myClg{}{}{\R^{n,n}}$
$$
x_i \lor x_j \equiv \myconjugate{\myconjugate{x}_i \myconjugate{x}_j} \to \Identity - (\Identity - x_i)(\Identity - x_j) = x_i + x_j - x_i x_j
$$
and it is simple to check that this expression is in $\mysetS$ since
$$
(x_i + x_j - x_i x_j) (x_i + x_j - x_i x_j) = x_i + x_j - x_i x_j
$$
with which the mapping of Boolean expressions of $n$ logical variables $x_i$ in the set of idempotents $\mysetS$ of $\myClg{}{}{\R^{n,n}}$ is complete and from now on we can treat any logical expression algebraically in Clifford algebra.

We support this statement with an example: it is simple to check that the logically equivalent relations $x_1 \lor x_2 \equiv \Identity - \myconjugate{x}_1 \myconjugate{x}_2$ are not algebraically equal in $\myAbelsa$ where they give respectively
$$
\begin{array}{l l l}
x_1 + x_2 & = & x_1 (x_2 + \myconjugate{x}_2) + (x_1 + \myconjugate{x}_1) x_2 = 2 x_1 x_2 + x_1 \myconjugate{x}_2 + \myconjugate{x}_1 x_2 \\
\Identity - \myconjugate{x}_1 \myconjugate{x}_2 & = & x_1 x_2 + x_1 \myconjugate{x}_2 + \myconjugate{x}_1 x_2
\end{array}
$$
whereas in $\mysetS$ the logical expression for $x_1 \lor x_2$, being $x_1 + x_2 - x_1 x_2 = x_1 x_2 + x_1 \myconjugate{x}_2 + \myconjugate{x}_1 x_2$, results algebraically equal to the expansion of $\Identity - \myconjugate{x}_1 \myconjugate{x}_2$.

\subsubsection*{Old parts of section~\ref{SAT_basics}}
A standard representation of Boolean associates $\{\mathrm{T}, \mathrm{F}\}$ to $\{1, 0\}$, logical OR to addition (modulo 2), logical AND to multiplication, and logical complement to the swap of $0$ and $1$, or equivalently, to the addition of $1$, then it is simple to prove that expanding expression (\ref{formula_SAT_std}) using the standard rules of arithmetic, together with $\literal_i \literal_i = \literal_i$ and $\literal_i \myconjugate{\literal}_i = 0$, a solution exists if and only if at least one of these terms is non zero.

\subsubsection*{Old parts of section~\ref{SAT_reformulation}}
Before proving this result we remark that when dealing with logical expressions we can attach two meanings to a product of literals: the first is an unique assignment of values to logical variables, for example we may interpret $\literal_i \literal_j \myconjugate{\literal}_k$ as $\literal_i \equiv \literal_j = \mathrm{T}, \literal_k = \mathrm{F}$ since only for these values the $1\SAT$ formula is $\mathrm{T}$. The second meaning is that of a $1\SAT$ formula with unassigned literals; let $S$ be this formula: we can then test if a given assignment of literals $S_0$ - in the first meaning - satisfies $S$ simply calculating $S \land S_0$: the result will be T if and only if $S_0$ satisfies $S$.

With our mapping in Clifford algebra this duality is respected and we can test if an assignment, \eg $\literal_i \literal_j \myconjugate{\literal}_k$, satisfies a given formula, \eg $\literal_i \myconjugate{\literal}_k$ simply calculating their product: if the result is not zero, namely $\mathrm{T}$, this means that the given assignment satisfies with the formula, in our example
$$
\literal_i \myconjugate{\literal}_k \; \; \literal_i \literal_j \myconjugate{\literal}_k = \literal_i \literal_j \myconjugate{\literal}_k \ne 0 \to \mathrm{T}
$$
in this sense we can read $\literal_i \myconjugate{\literal}_i = 0$ as the fact that $\literal_i = \mathrm{F}$ does not satisfy the formula $\literal_i$ and $\literal_i \literal_i = \literal_i$ as the fact that $\literal_i = \mathrm{T}$ satisfies formula $\literal_i$. With this in mind we proceed to prove the proposition.

\bigskip

Remembering the EFB rules
$$
p_i q_i \, p_i q_i = p_i q_i \quad p_i q_i \, q_i p_i = 0 \quad p_i q_i \, p_j q_j = p_j q_j \, p_i q_i \quad q_i p_i + p_i q_i = 1
$$
it is easy to see that to any logical OR one can assign the sum of the corresponding $\myClg{}{}{\R^{n,n}}$ elements \eg
$$
x_1 + \myconjugate{x}_3 \to q_1 p_1 + p_3 q_3
$$
and this formula is satisfied for \eg $x_1 = \mathrm{T}, x_3 = \mathrm{F}$, this is represented in $\myClg{}{}{\R^{n,n}}$ by the Clifford product
$$
(q_1 p_1 + p_3 q_3) q_1 p_1 p_3 q_3 = q_1 p_1 + p_3 q_3 \ne 0
$$
where we see that the ``substitution'' of a given assignment of variables ($q_1 p_1 p_3 q_3$) in a \SAT{} formula $(q_1 p_1 + p_3 q_3)$ is done with Clifford product and if the result is non zero the result is $\mathrm{T}$. It is easy to check the other three assignments that satisfy this formula and the only one that does not.

The procedure is the same for the logical AND that is associated to Clifford product
$$
x_1 \myconjugate{x}_2 \to q_1 p_1 p_2 q_2
$$
and also in this case it is easy to see that an assignment of variables satisfies the given formula ($x_1 \myconjugate{x}_2$) if and only if the Clifford product of the formula with the assignment is non zero. We see that an element like $q_1 p_1 p_2 q_2$ can be thought to represent both a formula and an assignment of variables and the assignment satisfies the formula if their product is non zero.

\bigskip

\noindent old comments following the general logical NOT formula (\ref{old_formula_complemented_assignement})

We remark that (\ref{old_formula_complemented_assignement}) needs a more formal proof since it uses the sum that we have seen can not be always mapped to a logical operation. We postpone a proof for a moment and use (\ref{old_formula_complemented_assignement}) and De Morgan's relations to get a more convenient form for a clause $C_j$....

\subsubsection*{Old parts of section~\ref{unSAT_algorithm}}
Before using these results for an algorithm that tests for unsatisfiability we need a technical
\begin{MS_lemma}
\label{technical_lemma}
given $s \in \mysetS - \{0\}$ then $s = \Identity$ if and only if, for any $1 \le i \le n$ then $s = \anticomm{q_i}{p_i} s'$ where $s' \in \mysetS$ and $\mygen_{2 i - 1} s' \mygen_{2 i - 1}^{-1} = s'$ and $\mygen_{2 i} s' \mygen_{2 i}^{-1} = s'$.
\end{MS_lemma}
\begin{proof}
The forward part of the lemma descends immediately from the expression of $\Identity$ in EFB (\ref{formula_identity_def}) and the observation that $\mygen_{j} \anticomm{q_i}{p_i} \mygen_{j}^{-1} = \anticomm{q_i}{p_i}$ for any $j$ since $\anticomm{q_i}{p_i} = 1$. Let us now suppose that for $s$ holds the given property: then for any $1 \le i \le n$ we may write $s = \anticomm{q_i}{p_i} s'$ and $\mygen_{2 i - 1} s \mygen_{2 i - 1}^{-1} = \mygen_{2 i} s \mygen_{2 i}^{-1} = s$ so that for all $1 \le i \le 2 n$ $\mygen_{i} s \mygen_{i}^{-1} = s$ and thus by \cite[Propostion~16.6]{Porteous_1995} $s = \delta \Identity$ with $\delta \in \R$ but since $s \in \mysetS - \{0\}$ the only possibility is $\delta = 1$.
\end{proof}

\begin{MS_Proposition}
\label{e_i_invariance_old}
Testing the invariance
$$
\mygen_{2 i - 1} \; \SAT \; \mygen_{2 i - 1} = \SAT
$$
for any single $1 \le i \le n$ is equivalent to test the equality of 2 derived $(k-1)\SAT$ problems in $n - 1$ logical variables $\SAT_1 = \SAT_2$ (defined in the sequel).
\end{MS_Proposition}

\begin{proof}
As already stated it is equivalent to test either logical equality of the logical expressions or the algebraic equality of the corresponding expressions in $\myClg{}{}{\R^{n,n}}$.

To test invariance of \SAT{} problem under any of the $n$ odd generators $\mygen_{2 i - 1}$ we start from the form (\ref{formula_SAT_EFB_3}) and we separate the $m$ clauses into three classes: those in which $\literal_i$ appears, those in which $\myconjugate{\literal}_i$ appears and the others in which neither appears. Obviously these three classes form three ``reduced'' $k\SAT$ problems that we call respectively $\SAT_1'$, $\SAT_2'$ and $\SAT_0$, clearly $\SAT = \SAT_0 \SAT_1' \SAT_2'$.

We start by $\SAT_0$: for all its clauses $\mygen_{2 i - 1} z_j \mygen_{2 i - 1} = z_j$ so they are invariant and $\SAT_0$ remains out of the game when testing $\mygen_{2 i - 1}$ invariance.

Let us now consider the product of the clauses of the first set $\SAT_1'$, those with $\literal_i$, that can be always put in the form (\ref{formula_SAT_EFB_3}). More precisely, since all these clauses contain $\literal_i$, when written in the complemented form $\Identity - z_j$ we are sure that $\myconjugate{\literal}_i$ appears in $z_j$ and so for these clauses
$$
\SAT_1' = \Identity - \myconjugate{\literal}_i \Delta_1
$$
with $\Delta_1$ ``free'' of literals $\literal_i$ and $\myconjugate{\literal}_i$. In a completely similar fashion $\SAT_2'$ takes the form
$$
\SAT_2' = \Identity - \literal_i \Delta_2 \dotinformula
$$
In summary we may write our original \SAT{}
$$
\SAT = \SAT_0 (\Identity - \myconjugate{\literal}_i \Delta_1) (\Identity - \literal_i \Delta_2) = \SAT_0 \left[\Identity - (\myconjugate{\literal}_i \Delta_1 + \literal_i \Delta_2)\right]
$$
and, as we have seen, $\SAT = 0$ is equivalent to $\Identity - \SAT = \Delta = \Identity$ so to apply lemma~\ref{technical_lemma} we consider the complementary expression
$$
\Identity - \SAT = \Identity - \SAT_0 \left[\Identity - (\myconjugate{\literal}_i \Delta_1 + \literal_i \Delta_2)\right]
$$
and since literals $\literal_i$ and $\myconjugate{\literal}_i$ do not appear neither in $\SAT_0$ nor in $\Delta_1$ nor in $\Delta_2$ but they implicitly appear, with all other anticommutators, in $\Identity$, we see that we can write this expression in the form
$$
\Identity - \SAT = (\myconjugate{\literal}_i + \literal_i) \left[\Identity - \SAT_0 (\Identity - \Delta_1)\right]
$$
\opt{margin_notes}{\mynote{mbh.note: note that the expression in square parenthesis is actually in Clifford algebra of $\R^{n-1,n-1}$}}%
if and only if $\Delta_1 = \Delta_2$ or, going to the complementary relation, if and only if
$$
\Identity - \Delta_1 = \Identity - \Delta_2 \dotinformula
$$
Clearly if this holds, we satisfy the conditions of technical lemma~\ref{technical_lemma} since $(\myconjugate{\literal}_i + \literal_i) = \anticomm{q_i}{p_i}$ and $\left[\Identity - \SAT_0 (\Identity - \Delta_1)\right]$ is free of literals $\literal_i$ and $\myconjugate{\literal}_i$, this means that $\Identity - \SAT$ is invariant for the reflections generated by $\mygen_{2 i - 1}$ and $\mygen_{2 i}$ and thus also \SAT.

It remains to show now that $\SAT_1 := \Identity - \Delta_1$ and $\SAT_2 := \Identity - \Delta_2$ are actually $(k-1)\SAT$ problems in $n - 1$ logical variables. Remembering the definition of \eg $\SAT_1'$ as the reduced problems containing only clauses in which $\literal_i$ appears explicitly, we showed easily that $\SAT_1' = \Identity - \myconjugate{\literal}_i \Delta_1$ with $\Delta_1$ free of $\literal_i$. It is a simple exercise to show that taking the very same clauses that form $\SAT_1'$ and \emph{removing} from them $\literal_i$ we obtain a $(k-1)\SAT$ problems in $n - 1$ logical variables whose form is precisely $\Identity - \Delta_1$.
\end{proof}

\bigskip

\noindent old --tentative-- final part of this paragraph

%\newpage
%\bigskip

Before tackling the resolution of this subproblem we remark:
\begin{itemize}
\item on average $\SAT_1$ and $\SAT_2$ are formed by $\frac{k}{2} \alpha$ clauses and $\SAT_0$ is formed by $(n - k) \alpha$ clauses;
\item always on average $\Delta_1$ and $\Delta_2$, when fully expanded, can contain at most $2^{\frac{k}{2} \alpha}$ terms, a number of terms that remains constant in the limit $n \to \infty$;
\item if $\SAT_1$ and $\SAT_2$ are formed by $p_{i,1}, p_{i,2}$ clauses each then $\sum_{i = 1}^n p_{i,1} + p_{i,2} = k m$ (assuming all clauses have exactly $k$ literals);
\item if our original \SAT{} problem had $k$ literals per clause in $n$ variables the generated expression $\Delta_1$ and $\Delta_2$ are formulations of $(k-1)\SAT$ with $n-1$ variables.

This can be verified with simple algebra manipulation: take the reduced $\SAT_1$problem made only by the clauses in which $\literal_i$ appears and then remove all the apearences of $\literal_i$, this produces a reduced $(k-1)\SAT$ with $n-1$ variables $\SAT_1'$, that put in its (\ref{formula_SAT_EFB_3}) form is exactly $\SAT_1' = \Identity - \Delta_1$;
\item to certify unsatisfiability we need to perform $n$ invariance test of the original problem but the first test that fails guarantees satisfiability but do not provide a solution. It is known that it is simple to use this algorithm to find an actual solution of the problem (test for unsatisfiability the two reduced problems with $\literal_i = \mathrm{T}, \mathrm{F}$ and then proceed to all other literals).
\item it remains to be established how hard is to test $\Delta_1 = \Delta_2$; if testing the logical equivalence is TAUT we know that this problem is substantially equivalent to \SAT{} but this new \SAT{} problem would be a 2\SAT....
\end{itemize}

\bigskip

\newpage
\noindent old --unfortunately wrong-- second part of paragraph 5

\begin{MS_Proposition}
\opt{margin_notes}{\mynote{mbh.note: there is a slightly different proof of this proposition in the last pages, actually it proved difficult}}%
\label{e_i_invariance_old2}
Testing the invariance
$$
\mygen_{2 i - 1} \; S \; \mygen_{2 i - 1} = S
$$
for any single $1 \le i \le n$ is equivalent to test the equality of 2 derived $(k-1)\SAT$ problems in $n - 1$ logical variables $S_1 = S_2$ (defined in the sequel).
\end{MS_Proposition}

\begin{proof}
As already stated testing the logical equivalence of two logical expressions is equivalent to testing algebraic equality of the corresponding expressions in $\mysetS$ of $\myClg{}{}{\R^{n,n}}$.

To test invariance of \SAT{} problem $S$ under any of the $n$ odd generators $\mygen_{2 i - 1}$ we start from its form (\ref{formula_SAT_EFB_3}) and we separate the $m$ clauses into three classes: those in which $\literal_i$ appears, those in which $\myconjugate{\literal}_i$ appears and the others in which neither appears. Obviously these three classes form three ``reduced'' $k\SAT{}$ problems that we call respectively $S_1'$, $S_2'$ and $S_0$, clearly $S = S_0 S_1' S_2'$.

We start by $S_0$: by (\ref{old_formula_literal_invariances}) for all its clauses $\mygen_{2 i - 1} z_j \mygen_{2 i - 1} = z_j$ so they are invariant and $S_0$ is out of the game when testing $\mygen_{2 i - 1}$ invariance.

Let us now consider the product of the clauses of the first set $S_1'$, those with $\literal_i$, that we put in form (\ref{formula_SAT_EFB_3}). More precisely, since all these clauses contain $\literal_i$, when written in the form $\Identity - z_j$ certainly $\myconjugate{\literal}_i$ appears in all $z_j$ and so for these clauses the form (\ref{formula_SAT_EFB_3}) is
$$
S_1' = \Identity - \myconjugate{\literal}_i \Delta_1
$$
with $\Delta_1$ ``free'' of literals $\myconjugate{\literal}_i$ and $\literal_i$. Similarly $S_2'$ takes the form
$$
S_2' = \Identity - \literal_i \Delta_2
$$
and we may write our original \SAT{} problem $S$
\begin{equation}
\label{formula_sat_subproblems_old}
S = S_0 (\Identity - \myconjugate{\literal}_i \Delta_1) (\Identity - \literal_i \Delta_2) = S_0 \left[\Identity - (\myconjugate{\literal}_i \Delta_1 + \literal_i \Delta_2)\right]
\end{equation}
and with (\ref{old_formula_literal_invariances})
$$
\mygen_{2 i - 1} S \mygen_{2 i - 1} = %\mygen_{2 i - 1} S_0 \mygen_{2 i - 1} \mygen_{2 i - 1} S_1 \mygen_{2 i - 1} \mygen_{2 i - 1} S_2 \mygen_{2 i - 1} =
S_0 \left[\Identity - (\myconjugate{\literal}_i \Delta_2 + \literal_i \Delta_1)\right]
$$
and we have thus to test wether
$$
S_0 \left[\Identity - (\myconjugate{\literal}_i \Delta_2 + \literal_i \Delta_1)\right] = S_0 \left[\Identity - (\myconjugate{\literal}_i \Delta_1 + \literal_i \Delta_2)\right]
$$
where both sides of the equality are elements of $\mysetS$. Since this is an algebraic equality we can rearrange it to get
%$$
%S_0 - \myconjugate{\literal}_i S_0 (\Delta_1 - \Delta_2) = S_0 - \literal_i S_0 (\Delta_2 - \Delta_1)
%$$
$$
S_0 \left[\Identity - \myconjugate{\literal}_i (\Delta_1 - \Delta_2)\right] = S_0 \left[\Identity - \literal_i (\Delta_1 - \Delta_2)\right]
$$
where $S_0, \Delta_1, \Delta_2$ are all without literals $\literal_i$ and $\myconjugate{\literal}_i$ and thus we may write $S_0 = (\literal_i + \myconjugate{\literal}_i) S_0$ that shows that there are two copies of $S_0$, one for each of the orthogonal subspaces $\literal_i$ and $\myconjugate{\literal}_i$ of $\myAbelsa$, see (\ref{formula_P_subspaces}). Since the factors in square parenthesis act selectively on only one of these subspaces the equality may hold if and only if $\Delta_1 = \Delta_2$ or, going to the complementary relation
$$
\Identity - \Delta_1 = \Identity - \Delta_2
$$
or $S_1 = S_2$ calling $S_l := \Identity - \Delta_l, l = 1,2$. It remains to show that $S_l$ are actually $(k-1)\SAT$ problems in $n - 1$ logical variables. Remembering the definition of \eg $S_1'$ as the reduced problems containing only clauses in which $\literal_i$ appears explicitly, we showed easily that $S_1' = \Identity - \myconjugate{\literal}_i \Delta_1$. It is a simple exercise to show that taking the very same clauses that form $S_1'$ and \emph{removing} from them $\literal_i$ we obtain a $(k-1)\SAT$ problem in $n - 1$ logical variables whose form (\ref{formula_SAT_EFB_3}) is $\Identity - \Delta_1$.
\end{proof}

Since the proposition holds for all $1 \le i \le n$ we have demonstrated:
\begin{MS_Corollary}
\label{unsatisfiability}
$S = 0$ if and only if the $n$ couples of derived $(k-1)\SAT$ problems in $n - 1$ logical variables of previous proposition are equivalent.
\end{MS_Corollary}

We can thus derive an algorithm that tests for unsatisfiability:
\begin{MS_Algorithm}
\label{unSAT_algorithm_outline}
\hfill
\begin{enumerate}
\item set $i = 1$;
\item generate the two auxiliary $(k-1)\SAT$ problems in $n - 1$ logical variables $S_1$ and $S_2$ relative to literal $\literal_i$;
\item if $S_1 \ne S_2$ the problem is satisfiable; STOP.
\item if $i < n$ set $i = i + 1$ and go to 2;
\item the problem is unsatisfiable; STOP.
\end{enumerate}
\end{MS_Algorithm}

The central point of this algorithm is to understand how difficult is to perform the test of step 3
\begin{equation}
\label{formula_sat_subproblem}
S_1 = S_2 \dotinformula
\end{equation}

Before tackling this problem we remark that if our original \SAT{} problem $S$ was a $2\SAT$ problem then the derived problems $S_1$ and $S_2$ generated in each test are $1\SAT$ problems that thus admit immediate solutions. In other words it is sufficient to compare the two resulting $1\SAT$ expressions, if they are equal the test is passed. Since, as already noted, the representation of any logical expression in $\mysetS$ is unique this comparison can be done in polynomial time%
\opt{margin_notes}{\mynote{mbh.note: dubious point, it could nevertheless be marginal}}%
{} and thus algorithm~\ref{unSAT_algorithm_outline} provides a simple explanation of the reason that makes $2\SAT$ problems simple to solve (polynomial).

\bigskip

We now analyze the complexity of the calculation of the test (\ref{formula_sat_subproblem}) for $3\SAT$ problems; in this case the two derived problems for each of the $n$ invariance tests are $2\SAT$ problems in $n - 1$ variables that, on average, are formed by $\frac{3}{2} \alpha$ clauses.

To test the logical equivalence $S_1 \equiv S_2$ we realize immediately that the two problems at hand can be: both unsatisfiable, and then the expressions are logically equivalent; one satisfiable and one unsatisfiable and then the expressions are logically not equivalent. It remains the case that both problems are satisfiable, in this case it is simple to prove that the expression are logically equivalent if and only if the two expressions $S_1 \myconjugate{S}_2$ and $\myconjugate{S}_1 S_2$ are both non satisfiable because it implies that the set of solutions of $S_1$ and of $S_2$ are equal and thus that $S_1 \equiv S_2$.

At a first look we may notice that in general $\myconjugate{S}_l = \Delta_l$ may contain an exponential number of terms that would make prohibitive the solution of the new derived problems $S_1 \myconjugate{S}_2$ and $\myconjugate{S}_1 S_2$. But from the logical wiewpoint we may notice that $\myconjugate{S}_l \equiv z_1 \lor z_2 \lor \cdots \lor z_p$ where, as in (\ref{formula_SAT_EFB_2}), $z_j$ is the product of the complemented literals of the $p$ clauses of the problem at hand. In this case we can easily prove the technical
\begin{MS_lemma}
\label{2SAT_equivalence}
Given two satisfiable formulas $S_l$ and $S_k = \prod_{j = 1}^p (\Identity - z_j)$ the formula $S_l \myconjugate{S}_k$ is unsatisfiable if and only if the $p$ subproblems $S_l z_j$ for $j = 1, \ldots, p$ are all unsatisfiable.
\end{MS_lemma}
\begin{proof}
By logical properties
$$
S_l \myconjugate{S}_k \equiv S_l (z_1 \lor z_2 \lor \cdots \lor z_p) \equiv S_l z_1 \lor S_l z_2 \lor \cdots \lor S_l z_p
$$
it is clear that $S_l z_j \in \mysetS$ and by the remarks following proposition~\ref{S_properties} $S_l z_j$ is just problem $S_l$ with the literals of $z_j$ given their values and thus a new, simpler, \SAT{} problem. $S_l \myconjugate{S}_k$ is thus unsatisfiable if and only if all the $p$ reduced problems $S_l z_j$ are unsatisfiable.
\end{proof}

We can exploit these results to outline an algorithm that tests logical equivalence (\ref{formula_sat_subproblem}) of two $2\SAT$ problems:
\begin{MS_Algorithm}
\label{SAT_equivalence_algorithm_outline}
\hfill
\begin{enumerate}
\item solve $S_1$ and $S_2$;
\item if one is unsatisfiable and the other is satisfiable the problems are not equivalent; STOP.
\item if the problems are both unsatisfiable the problems are equivalent; STOP.
\item generate the $p_1$ expressions $z_j$ for $S_1$ and the $p_2$ $z_j'$ for $S_2$;
\item solve in succession the $p_1 + p_2$ reduced $2\SAT$ problems of the form $S_1 z_j'$ and $S_2 z_j$, if any is satisfiable the problems are not equivalent; STOP.
\item the problems are equivalent; STOP.
\end{enumerate}
\end{MS_Algorithm}

Supposing that the solution of $2\SAT$ problems is \bigO{n} and since the total number of clauses of $S_1$ and $S_2$ problems, $p_1 + p_2$, is \bigO{n-1} the total running time of algorithm~\ref{SAT_equivalence_algorithm_outline} is \bigO{n^2}.

We can thus conclude that the total running time of algorithms~\ref{unSAT_algorithm_outline} and \ref{SAT_equivalence_algorithm_outline} combined for $3\SAT$ problems is $\bigO{n^3}$.

Clearly this algorithm for $3\SAT$ could be used to test step 3 of algorithm~\ref{unSAT_algorithm_outline} for a $4\SAT$ problem that would thus result $\bigO{n^5}$. This can be generalized to a $\bigO{n^{2 k - 3}}$ algorithm for $k\SAT$ problems for $k > 1$.

} % note finali: stampate solo se all'inizio c'è l'opzione final_notes

\end{document}